\documentclass[10pt]{article}

\usepackage[T1]{fontenc}
\usepackage{amscd, amsfonts, amsmath, amssymb, amstext, amsthm, caption, epsfig, fancyhdr, float, graphicx, latexsym, mathtools, multicol, multirow, algorithm, chngcntr}
\usepackage{kpfonts}
\usepackage[lofdepth,lotdepth]{subfig}
\usepackage{authblk}
\usepackage[utf8]{inputenc}
\usepackage{bm} 
\usepackage{enumerate}
\usepackage{relsize}
\usepackage{booktabs}
\usepackage[numbers]{natbib}
\usepackage{color}
\usepackage{bm}
\usepackage{parskip}
\usepackage{newfloat}
\usepackage{listings}
\usepackage{xcolor}
\usepackage{eurosym}

\DeclareFloatingEnvironment[fileext=frm,placement={!ht},name=Code Fragment]{codefragment}

\lstdefinelanguage{Solidity}{
    keywords={pragma, solidity, contract, returns, memory, public, pure, mapping, struct},
    morekeywords=[2]{uint8, uint256, uint16, address, bytes32},
    morekeywords=[3]{function},
    morekeywords=[4]{require},
    sensitive=true,
    comment=[l]{//},
    morecomment=[s]{/*}{*/},
    morestring=[b]",
}

\usepackage[pdfpagelabels,colorlinks=true,linkcolor=blue,anchorcolor=blue,citecolor=blue,filecolor=blue,menucolor=blue,runcolor=blue,urlcolor=blue]{hyperref}

\usepackage{algpseudocode} 

\usepackage{pgf}
\usepackage[utf8]{inputenc}\DeclareUnicodeCharacter{2212}{-}

\usepackage{caption}
\usepackage[capitalise,nameinlink,noabbrev]{cleveref}
\crefname{equation}{}{}
\crefname{enumi}{}{}

\graphicspath{{Figures/}}

\makeatletter
\newcounter{algorithmicH}
\let\oldalgorithmic\algorithmic
\renewcommand{\algorithmic}{%
  \stepcounter{algorithmicH}
  \oldalgorithmic}
\renewcommand{\theHALG@line}{ALG@line.\thealgorithmicH.\arabic{ALG@line}}
\makeatother

\makeatletter
\DeclareCaptionLabelFormat{andtable}{#1~#2  \&  \tablename~\thetable}
\makeatletter

\makeatletter
\DeclareCaptionLabelFormat{andtable}{#1~#2  \&  \tablename~\bm{\theta}ble}
\makeatletter

\newtheorem{remark}{Remark}[section]
\newtheorem{ex}{Example}

\newcommand*{\GammaDist}{\mathsf{Gamma}}

\newcommand*{\NormalDist}{\mathsf{Normal}}

\newcommand*{\PoissonDist}{\mathsf{Poisson}}

\usepackage{xspace}

\newcommand*{\iid}{\textsc{iid}\@\xspace}

\definecolor{MyBlue}{HTML}{1f77b4}
\definecolor{MyGreen}{HTML}{2ca02c}
\definecolor{MyRed}{HTML}{d62728}
\definecolor{MyPurple}{HTML}{9467bd}
\definecolor{MyBrown}{HTML}{8c564b}

\usepackage[percent]{overpic}

\usepackage{float}
\usepackage{microtype}

\renewcommand{\tilde}{\widetilde}
\renewcommand{\hat}{\widehat}

\usepackage{cancel}

\usepackage[marginratio=1:1,height=584pt,width=480pt,tmargin=90pt]{geometry}

\usepackage{tikz}
\usetikzlibrary{positioning, arrows.meta}

\begin{document}

\title{Collaborative and parametric insurance on the Ethereum blockchain}
\author[1]{Pierre-Olivier Goffard \footnote{Email: \href{mailto:goffard@unistra.fr}{goffard@unistra.fr}.}}
\affil[1]{\footnotesize Université de Strasbourg, Institut de Recherche Mathématique Avancée, Strasbourg, France}
\author[2]{Stéphane Loisel\footnote{Email: \href{mailto:stephane.loisel@lecnam.net}{stephane.loisel@lecnam.net}.}}
\affil[2]{\footnotesize CNAM, Efab, Lirsa, Paris, France}

\maketitle
\vspace{3mm}

\begin{abstract}
This paper introduces a blockchain-based insurance scheme that integrates parametric and collaborative elements. A pool of investors, referred to as surplus providers, locks funds in a smart contract, enabling blockchain users to underwrite parametric insurance contracts. These contracts automatically trigger compensation when predefined conditions are met. The collaborative aspect is embodied in the generation of tokens, which are distributed to surplus providers. These tokens represent each participant's share of the surplus and grant voting rights for management decisions. The smart contract is developed in Solidity, a high-level programming language for the Ethereum blockchain, and deployed on the Sepolia testnet, with data processing and analysis conducted using Python. In addition, open-source code is provided and main research challenges are identified, so that further research can be carried out to overcome limitations of this first proof of concept. 
\end{abstract}

\emph{MSC 2010}: 91B30.\\
\emph{Keywords}: Blockchain, parametric insurance, actuarial science.

\tableofcontents
\section{Introduction} \label{sec:intro}

 A blockchain is a decentralized ledger maintained by consensus within a peer-to-peer network. The nodes in this network agree on a shared history of data by executing an algorithm known as a consensus protocol. The Bitcoin blockchain, as described by \citet{Nakamoto2008}, simply records transactions between bitcoin users. In contrast, the Ethereum blockchain, introduced by \citet{Buterin2014}, is also equipped with a compiler to execute pieces of code called smart contract. Ethereum is designed to operate as a "world computer," enabling developers to deploy decentralized applications (dApps) that users can interact with directly.

Through these decentralized applications, Ethereum extends the capabilities of blockchains beyond simple fund transfers, enabling more complex financial operations and fostering the emergence of a new financial ecosystem known as decentralized finance (DeFi), as described by \citet{Schaer2020}. Traditional finance relies on trusted third parties and centralized databases, which involve intermediary costs, lack transparency, operate only during limited hours, and are susceptible to cyberattacks. In contrast, distributed ledgers are publicly accessible, replicated in multiple locations, and function continuously, 24/7. Furthermore, the code that governs decentralized applications is auditable, improving transparency. For further discussion, we refer the reader to \citet[Chapter 2]{Lipton2021}.

Some criticisms of the financial system can also be extended to the insurance sector. Insurance policies are often complex and difficult for the general public to understand. Premium calculations are typically based on historical data that are privately held by insurance companies, which limits transparency. In addition, the claim settlement process is lengthy and costly as it requires on-site assessments by one or more experts.

One response to these challenges is parametric insurance, also known as index-based insurance. Parametric insurance provides compensation based on predefined triggers, such as specific weather conditions, rather than relying on traditional, often time-consuming, claims assessments. The main advantages include fast payouts and reduced management costs. When the verification of the triggering condition is done using a reliable and independent data source, it increases policyholders' confidence in the system.

Parametric insurance is not a new concept. It emerged in Asia during the 1990s, allowing farmers to buy protection against income loss due to poor harvests following adverse weather conditions. Since then, parametric insurance has been used to cover perils like natural catastrophes. Notable examples include the Caribbean Catastroph Risk Insurance Facility (\href{https://www.ccrif.org/}{CCRIF}), launched in 2007, and the African Risk Capacity (\href{https://www.arc.int/}{ARC}) founded in 2014. Fast payouts in the context of natural disaster relief are crucial, as they enable the rapid deployment of resources by government agencies and non-governmental organizations providing immediate assistance.

In recent years, parametric insurance schemes have emerged in mass-market products, such as flight delay insurance offered by China Southern Airlines, which provides payouts of up to 300 yuan (approximately US\$40) for delays of three hours or more. Note that death benefits in traditional insurance often consist of a fixed amount that varies according to limited number of parameters, such as the number of children or marital status. This forms a prominent example of a parametric insurance contract that is widely accepted and commercialized.  

The main downside of parametric insurance is basis risk, where the insured's losses may not match the coverage amount, or the parameter may not trigger. Effective structuring and pricing require understanding the policyholder's exact risks and choosing the right parameter. The availability of finer granularity data and advances in predictive modeling can mitigate basis risk. Hybrid insurance solutions that combine parametric and standard indemnity mechanisms can also be effective, see for instance \citet{Lopez2025}. An initial payment following the parametric trigger can provide immediate relief, and this amount can later be deducted from the compensation resulting from the traditional claims assessment process. The policyholder has the option to either maintain parametric insurance or purchase an additional traditional insurance contract to cover basis risk.  

For a historical overview of parametric insurance, we refer the reader to the press article by \citet{Clere2022} in InsurTech Digital magazine. Interesting discussions on the use of parametric insurance to address climate risk issues can be found on the websites of the World Economic Forum \cite{WEF} and the National Association of Insurance Commissioners \cite{NAIC}. The application of parametric insurance to crop insurance has been discussed in the literature (see \citet{Abdi2022} and \citet{Conradt2015}). The simplicity of the insurance mechanism makes it suitable for encoding in a smart contract. A blockchain can store data immutably and transparently, making the information associated with the triggering event and compensation available to all interested parties. Connecting the blockchain to users' wallets allows for fast and automated processing of indemnity payments.

We present a smart contract on the Ethereum blockchain to manage a portfolio of parametric insurance policies. We have developed an actuarial framework suited for weather-linked triggering events, specifically monitoring daily rainfall. This framework establishes ground rules for underwriting, rate-making, and managing risk. Our smart contract is designed to operate with minimal human interference, maximizing automation. Every interaction with the smart contract incurs a cost, as any transaction submitted to the blockchain requires a fee to be processed by the network of nodes. Therefore, optimizing the number of transactions is crucial for reducing operational costs. Actuarial methods for calculating risk capital are often too complex to be directly implemented in a high-level language like Solidity, which, for instance, only supports integer-based arithmetic. Additionally, deploying a smart contract on the blockchain entails costs that grow with its complexity, emphasizing the need for simplicity in design. One of our key contributions is the simplification of these computations through justified approximations. In particular, we propose a straightforward recursive formula to determine the solvency capital required to adequately cover the risks of a parametric insurance portfolio.
 
The smart contract structure is inspired by the discussion in \citet{Cousaert2022}, from which we adopt the role distribution and the  "tokenization" mechanism of risk. The smart contract owner writes the code and sets the initial values for parameters such as the premium loading and the risk level required for risk capital allocation. Investors participate by locking funds in the smart contract, in exchange for which they receive protocol tokens. These funds are denominated in \(\text{ETH}\), the native cryptocurrency of the Ethereum blockchain. The contribution of each investor is reflected in their token holdings. The value of the token, or equivalently its exchange rate against \(\text{ETH}\), fluctuates over time depending on whether compensation are being paid to policyholders. Conversely, the token value increases if no compensation is paid at settlement. Token ownership enables governance of the insurance protocol. The contract owner may update the smart contract parameters based on recommendations from a board composed of token holders. This governance model transforms the smart contract into a Decentralized Autonomous Organization (DAO). The tokenization of risk on the blockchain can be viewed as a novel way to design insurance-linked securities, as outlined in \citet{Barrieu2012}. By enabling policyholders to become investors, our smart contract creates a collaborative insurance scheme, akin to those described in \citet{Feng2023}. Investors, referred to as Surplus Providers (SPs), are analogous to liquidity providers in decentralized exchanges, which serve as blockchain-based trading venues for crypto assets; see, for instance, \citet{Mohan2022}. 

Future policyholders interact with the protocol to seek protection against adverse events. An insurance agreement can only be concluded if the funds provided by the investors are sufficient to cover the risk. The smart contract funds must remain above a specified threshold at all times, as prescribed by standard insurance regulations such as the European directive Solvency II.

We showcase the concrete implementation of our framework using the Solidity programming language. This includes instructions on deploying the contract on the Ethereum testnet, interacting with it, and retrieving data via the \href{https://sepolia.etherscan.io/}{Etherscan} Application Programming Interface (API). For a comprehensive overview of how the Ethereum blockchain operates and a tutorial on coding in Solidity, we refer the reader to the book of \citet{Antonopoulos}. 

This paper aims to show the feasibility of deploying a beta version of such a smart contract, where certain complexities are intentionally simplified. We also identify the main limitations of this simplified framework and provide the research and actuarial communities with an open-source beta version of the smart contract. This prototype serves as a foundation for developing generalizations and improved versions that address these limitations.

 The rest of the paper is organized as follows. \cref{sec:blockchain_technology} provides a brief overview on blockchain technology. \cref{sec:parametric_insurance} presents an actuarial framework for managing parametric insurance contract portfolios. \cref{sec:smart_conract_for_parametric_insurance} focuses on the design of the smart contract. We introduce the participants, including the contract owner, surplus providers, and policyholders, in \cref{ssec:roles}. A smart contract is characterized by state variables, which we describe in \cref{ssec:state_variable}. Examples of state variables include the smart contract’s balance and the tokens held by each participant. Users interact with the smart contract by invoking functions, and the possible actions and their impacts on the system’s state are detailed in \cref{ssec:smart_contract_methods}. \cref{sec:smart_contract_interaction} demonstrates how the smart contract operates through an event-driven scenario, showcasing all its functions and culminating in the liquidation and reset of the smart contract. The main limitations of this beta version of the smart contract are discussed in \cref{sec:limitations}, along with potential ideas for addressing them.

\section{Blockchain technology and smart contracts}\label{sec:blockchain_technology}


The concept of blockchain technology was first introduced with Bitcoin, a decentralized digital currency launched in 2008 with \citet{Nakamoto2008}. Bitcoin's blockchain is a public ledger that records all transactions in a secure, immutable, and transparent manner. Its core innovation lies in achieving consensus among distributed nodes without the need for a central authority, using a Proof-of-Work (PoW) mechanism.

However, the Bitcoin blockchain was designed with a narrow scope: it enables peer-to-peer transfer of value but offers very limited programmability. As a result, more complex applications, such as decentralized financial instruments or autonomous organizations, are not feasible on the Bitcoin network. To overcome these limitations, the concept of \emph{programmable blockchains} emerged, most notably with the introduction of Ethereum in 2015, see \citet{Buterin2014}. Ethereum extends the blockchain model by embedding a \emph{Turing-complete virtual machine}, the Ethereum Virtual Machine (EVM), which enables the execution of code on-chain. This innovation allows developers to write \emph{smart contracts}—autonomous programs that run exactly as programmed—opening the door to a wide range of decentralized applications (dApps) across finance, insurance, governance or digital ownership, some concrete projects are mentionned in \cref{tab:dapp-examples}.
\begin{table}[h!]
\scriptsize
\centering
\begin{tabular}{llp{6.5cm}l}
\toprule
\textbf{Category} & \textbf{Project} & \textbf{Description} & \textbf{URL} \\
\midrule
Finance (DeFi) & Uniswap & A decentralized exchange (DEX) that allows users to swap ERC-20 tokens using liquidity pools and an automated market maker model. & \url{https://uniswap.org} \\
\midrule
Insurance & Etherisc & Provides decentralized parametric insurance, e.g., crop insurance for farmers, using smart contracts and weather data oracles. & \url{https://etherisc.com} \\
\midrule
Governance & Aragon & A platform for building and managing decentralized autonomous organizations (DAOs) with voting and fund management capabilities. & \url{https://aragon.org} \\
\midrule

Digital Ownership & Decentraland & A virtual reality platform on Ethereum where users can buy, develop, and monetize virtual plot of land using NFTs (Non-Fungible Tokens). & \url{https://decentraland.org} \\
\bottomrule
\end{tabular}
\caption{Examples of decentralized applications across sectors.}
\label{tab:dapp-examples}
\end{table}
"ERC-20 tokens" refer to fungible tokens on the Ethereum blockchain that are interchangeable and used for various transactions and financial operations, while "NFTs"  represent unique digital assets, such as collectibles or in-game items, that are not interchangeable and verify ownership on the blockchain. Token generation is a central feature of the Ethereum blockchain as it enables the creation of diverse digital assets that power decentralized applications and facilitate transactions between users.

Ethereum is currently the most widely adopted programmable blockchain, and as such was selected for the implementation of our parametric insurance framework. Ethereum benefits from the largest and most active developer community among blockchain platforms. This has led to a wealth of tooling, documentation, libraries (e.g., \href{https://hardhat.org/}{Hardhat}, \href{https://archive.trufflesuite.com/}{Truffle}, \texttt{web3.js}), and community support, which greatly facilitate the development and deployment of decentralized applications. Many formal verification tools and auditing services are tailored for Ethereum-based contracts, increasing trust and reducing the likelihood of critical vulnerabilities. The Ethereum blockchains successfully transited from Proof-of-Work to Proof-of-Stake inducing a significant reduction of its energy consumption and improved network scalability, helping to ensure long-term sustainability and compliance with environmental standards. Lastly, Ethereum hosts the majority of DeFi protocols, NFT platforms, and DAO frameworks. Its large user base and vibrant application ecosystem make it the natural choice for projects that require integration with or exposure to these decentralized systems.  We acknowledge the existence of other programmable blockchains that offer valuable features. \href{https://www.binance.org/en/smartChain}{Binance Smart Chain (BSC)} offers compatibility with Ethereum’s EVM but is more centralized, with fewer validators, potentially limiting trust and decentralization. \href{https://solana.com/}{Solana} provides high throughput and low latency but suffers from frequent network outages and a more complex programming environment. \href{https://www.avax.network/}{Avalanche} supports custom blockchains and fast finality, but its developer ecosystem and tooling are less mature compared to Ethereum.

\section{Parametric insurance portfolio management framework}\label{sec:parametric_insurance}
This section presents an actuarial framework for parametric insurance tailored to an implementation on the Ethereum blockchain. While this framework supports our analysis, we emphasize that it is not the primary contribution of our work. We further acknowledge that it could be extended or refined to accommodate a broader class of insurance solutions.

\subsection{Parametric insurance contract over time}\label{ssec:parametric_inusrance_contract}

A parametric insurance contract pays a predefined compensation \( l \) if some observable quantity \((Q_t)_{t\geq 0}\) reaches a threshold \(\overline{Q}\) at some time \( T \). The quantity \( Q := (Q_t)_{t \geq 0} \) is a stochastic process defined on a filtered probability space \((\Omega, \mathcal{F}, (\mathcal{F}_t)_{t \geq 0}, \mathbb{P})\). A compensation is paid if the \(\mathcal{F}_T\)-measurable event: \(\{Q_T > \overline{Q}\}\) occurs. A parametric insurance contract is represented as a 5-tuple \((S, T, Q, \overline{Q}, l)\), where \( S \) denotes the time at which the contract is underwritten. We outline in \cref{rem:interval_instead_of_punctual} and \cref{rem:index_based_compensation} two extensions associated to the time of the triggering event and the amount compensated respectively.

\begin{remark}\label{rem:interval_instead_of_punctual}
When the trigger is based on weather condition, the insurance coverage usually concerns a time interval. The triggering event definition must be suited to the meteorological quantity we monitor. We consider daily rainfalls as an example later in the article and so a typical event would be
\[
\text{"The cumulative precipitation on the 1st of May, 2025 in Strasbourg, France, exceeded }10\text{mm"}.
\]
We therefore consider $Q$ as a discrete time stochastic process or a time serie corresponding to the amount rain that fell on a specific day. A continuous time model would be better suited for other situations like flood risk for example. The trigger would be based on the cumulated rain over a rolling window of several days.  If we consider windspeed then time \( T \) should be replaced by a time interval \((T_{-}, T_{+})\) and the triggering event then becomes
\[
\left\{\underset{t \in (T_{-}, T_{+})}{\max}Q_t > \overline{Q}\right\}, \text{ or } \left\{\underset{t \in (T_{-}, T_{+})}{\min} Q_t > \overline{Q}\right\},
\]
where $Q$ is a continuous time stochastic process. Our approach could be extended to accomodate such refinements.
\end{remark}

\begin{remark}\label{rem:index_based_compensation}
The compensation can be a function \( l: Q_T \mapsto l(Q_T) \) of the quantity \( Q_T \). Here, we only consider a pre-specified lump sum payment.
\end{remark}
Define the status of the contract over time as:
\begin{equation}\label{eq:contract_status}
\xi_t = 
\begin{cases} 
0, & \text{if } t < T , \\
1, & \text{if } t \geq T \text{ and } Q_T \leq \overline{Q} , \\
2, & \text{if } t \geq T \text{ and } Q_T > \overline{Q} . \\
\end{cases}
\end{equation}

The state space of \((\xi_t)_{t \geq 0}\) is \(\mathcal{S} = \{0, 1, 2\}\). Each state corresponds to a specific condition of the contract:

\[
0 = \text{"open"}, \quad 1 = \text{"closed without compensation"}, \text{ and} \quad 2 = \text{"closed with compensation"}.
\]
The smart contract may be forced to file for bankruptcy if the surplus falls below the minimum capital requirement at any point between \( S \) and \( T \). If this occurs, the remaining active policies are cancelled, and the premiums are refunded to the policyholders. This possibility implies to add a fourth state 
\[
3 = \text{"cancelled"},
\]
which is covered in \cref{ssec:settlement}.

The pure premium associated with a contract \((S, T, Q, \overline{Q}, l)\) is given by 
 \[ \mathbb{P}(Q_T > \overline{Q}|\mathcal{F}_S)\cdot l.\] 
 The probability of the event \(\{Q_T > \overline{Q}\}\) estimated at some time \(t\in[0,T)\) is a stochastic process defined by 
 \begin{equation}\label{eq:theta_stochastic_process}
 \theta_t = \mathbb{P}(Q_T > \overline{Q}|\mathcal{F}_t)\text{ for  }t\in[0,T).
 \end{equation}
 The filtration \((\mathcal{F}_t)_{t\geq 0}\) gathers all the available information at time \(t \leq T\) relevant to the occurrence of event \(\{Q_T > \overline{Q}\}\) or equivalently the study of the quantity \((Q_t)_{t\geq 0}\). For weather-linked events, the quantity \(Q_T\) may be forecasted based on the available information at time \(t < T\) using weather forecast models. Concretely, the filtration \((\mathcal{F}_t)_{t\geq 0}\) contains both observations and predictions available at time \(t \geq 0\) and this must be accounted for in our actuarial pricing framework. Constraints on \(S\) can be imposed to mitigate the risk of adverse selection as outlined in \cref{rem:adverse_selection_linked_to_prediction}.

\begin{remark}\label{rem:adverse_selection_linked_to_prediction}
From a practical standpoint, we can assume the existence of a time \( S^\ast = T - \Delta < T \) before which the triggering event cannot be reliably anticipated. Therefore, the estimation of the probability of \(\{Q_T > \overline{Q}\}\) relies solely on historical, observed data. If policyholders have access to a more accurate predictive model than that of the insurers, it could lead to adverse selection. A straightforward approach to mitigate this risk is to stipulate that a policy \((S, T, Q, \overline{Q}, l)\) cannot be underwritten if \( S > S^\ast \). We can assume that:

\[
 \mathbb{P}(Q_T > \overline{Q} | \mathcal{F}_s) = \theta, \text{ for any } s \leq S^\ast.
\]
\end{remark}

We consider as a use case a parametric insurance coverage against rain episodes. An illustration is provided in \cref{ex:parametric_insurance_against_rain} with a concrete estimation of the probability \(\theta\).
\begin{ex}\label{ex:parametric_insurance_against_rain}
Consider the insurance policy \((S, T, Q, \overline{Q}, l)\), where \(Q = (Q_t)_{t \geq 0}\) is the daily precipitation height measured at the meteo station of Strasbourg airport (Entzheim). We assume that \(S \leq S^*\). We consider a parametric model for the daily precipitation height where \(Q_T|\mathcal{F}_S\) is distributed as a compound Poisson-gamma random variable. We have
\[
Q_T|\mathcal{F}_S \overset{\mathcal{D}}{=} \sum_{k = 1}^{N_T} U_{k,T},
\]
where \(N_T\) is a Poisson random variable \(\PoissonDist(\lambda_T)\) and the \(U_k\) are \iid gamma random variables \(\GammaDist(\alpha_T, \beta_T)\) independent from \(N_T\). This model was proposed by \citet{Dunn2004} with \(N_T\) corresponding to the number of rain episodes during a day and the \(U_k\)'s characterize their intensity. We assume that the parameters of the model \(\lambda_T\), \(\alpha_T\) and \(\beta_T\) simply depend on the month associated with the date \(T\). We calibrate this model using daily data from the years \(2000\) to \(2023\) provided by Météo France and retrieved from the \href{https://www.data.gouv.fr/fr/datasets/donnees-climatologiques-de-base-quotidiennes/}{data.gouv.fr}. The time series is shown in \cref{fig:ts_rainfall_strasbourg}.

\begin{figure}[!ht]
\centering
  \includegraphics[width=0.6\linewidth]{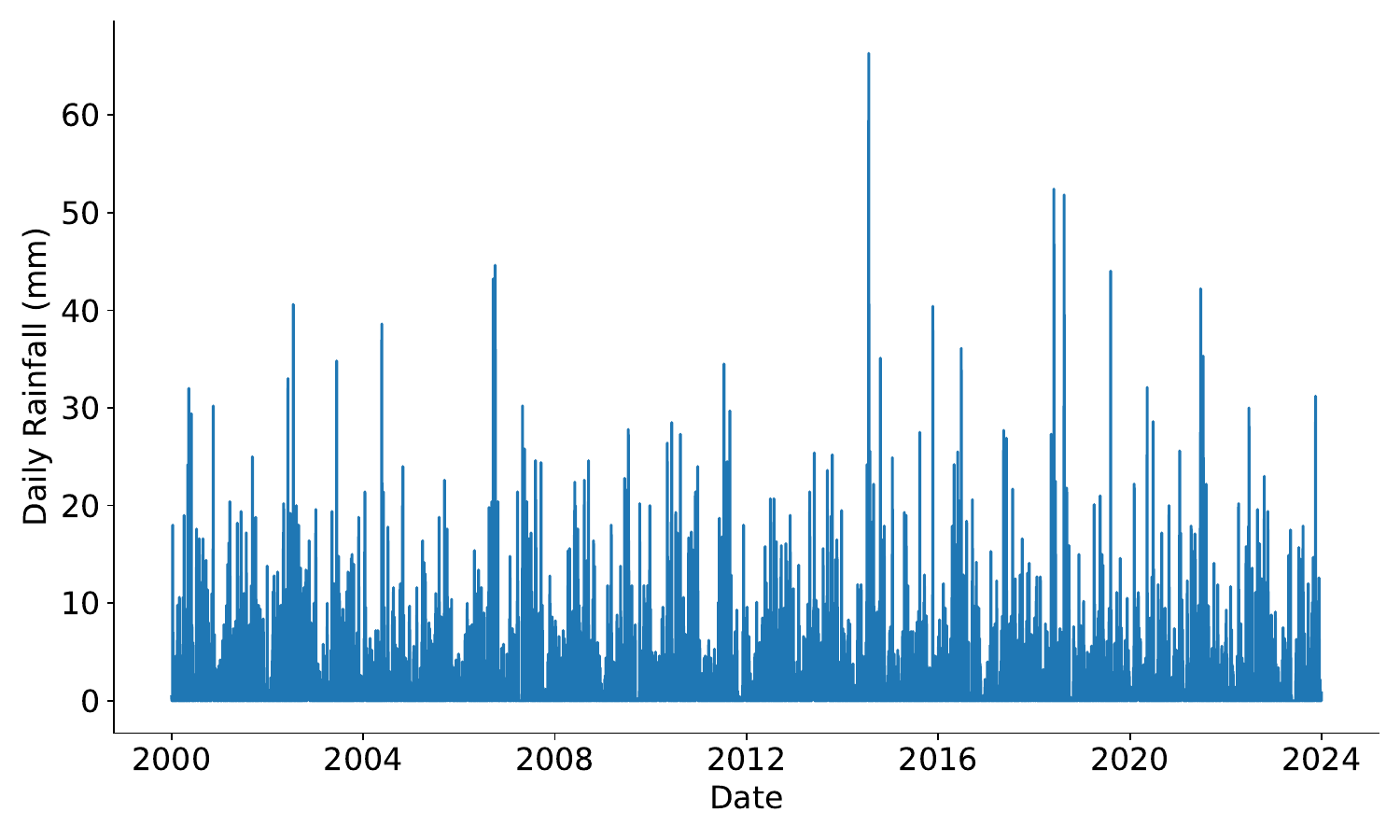}
  \caption{Daily rainfall measurements recorded at the STRASBOURG-ENTZHEIM station, spanning from January 1, 2000, to December 31, 2023.}
  \label{fig:ts_rainfall_strasbourg}
\end{figure}

The model is fitted using a simple partial method of moments so that
\(\hat{\lambda} = -\log(\hat{p}_0)\), where \(\hat{p}_0\) is the proportion of days without rain episodes over a month. We further have
\[
\hat{\alpha} = \frac{\mu^2}{\hat{\lambda}\sigma^2 - \mu^2} \text{ and } \hat{\beta} = \frac{\mu}{\hat{\lambda}\hat{\alpha}},
\]
 where $\mu$ and $\sigma$ denote the empirical mean and variance of the daily precipitations for rainy days. This simple method is described for instance in \citet{Goffard2022}. We have a compound Poisson-Gamma model for the STRASBOURG-ENTZHEIM station for each month; the estimations of the parameters allow us to estimate the number of rain episodes and their intensity for each month as shown in \cref{fig:rainy_days_and_rain_episode_intensity_strasbourg_by_month}.

\begin{figure}[!ht]
\centering
\includegraphics[width=0.6\linewidth]{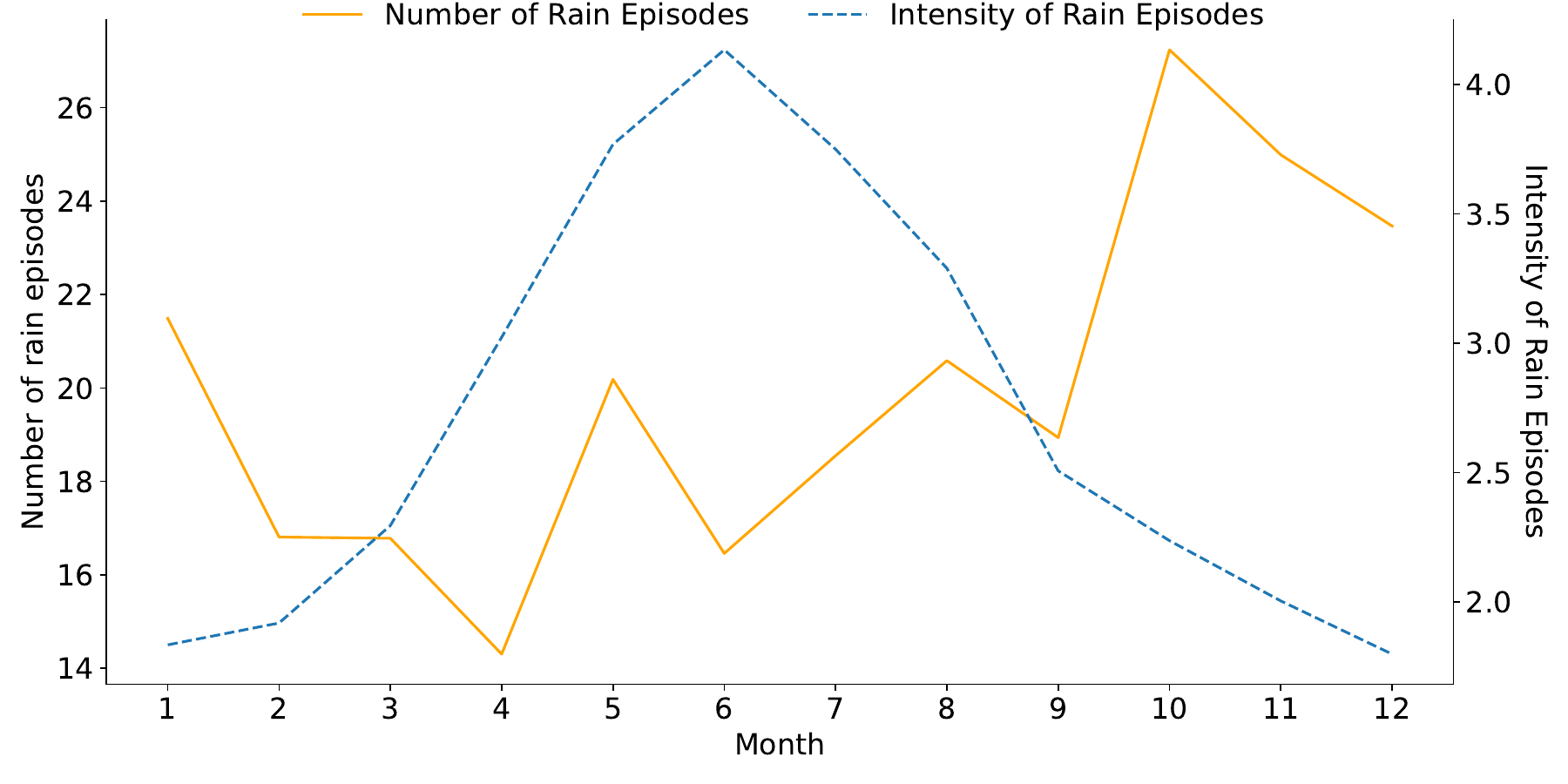}
\caption{Number of rain episodes per month and intensity of the rain episodes depending on the month in Strasbourg}
\label{fig:rainy_days_and_rain_episode_intensity_strasbourg_by_month}
\end{figure}

The climate in Strasbourg is "semi-continental". In terms of precipitation, it is characterized by low precipitation levels during the winter months (replaced by snow) and few but intense rain episodes during the summer months. The model then provides us with the probability \(\theta\) associated with the threshold \(\overline{Q}\) for each month, see \cref{fig:probability_strasbourg_by_month}.

\begin{figure}[!ht]
\centering
\includegraphics[width=0.6\linewidth]{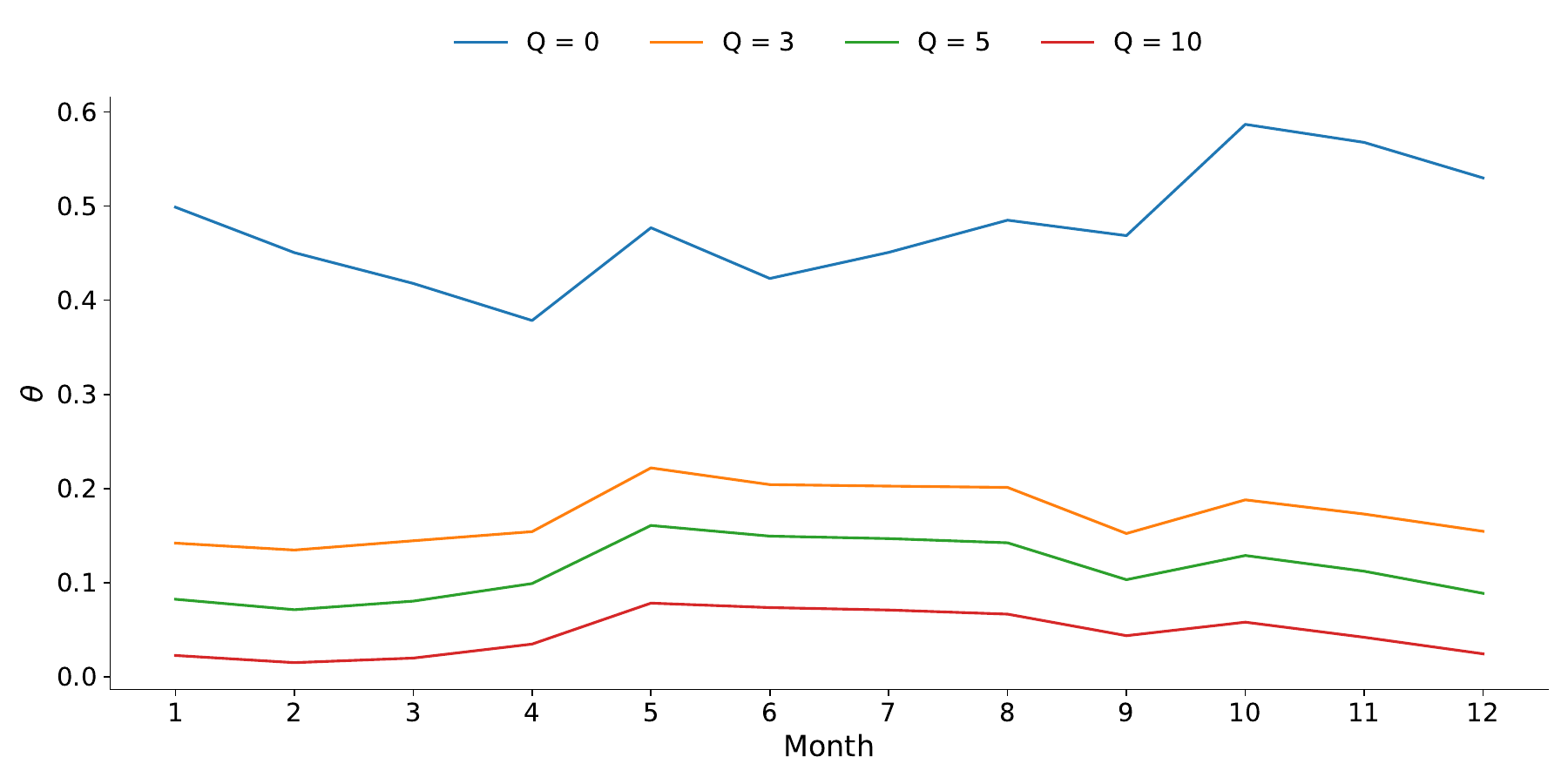}
\caption{Estimation of the probability \(\mathbb{P}(Q_T > \overline{Q})\) depending on the month and the threshold \(\overline{Q} \in \{0, 3, 5, 10\}\)}
\label{fig:probability_strasbourg_by_month}
\end{figure}

The estimation of \(\theta\) yields the pure premium of the contract as \(\theta \cdot l\).

\end{ex}

\begin{remark}\label{rem:polynomial_approximation_pricing_formula}
The pricing of our parametric insurance contract relies on a piecewise constant function \( T \mapsto \theta(T) \). From a practical perspective, dealing with dates in Solidity (the programming language of the Ethereum blockchain) is not straightforward; furthermore, having to handle 12 clusters of dates is tedious. Our solution consists of writing \( T \) as an integer number between 1 and 365 and approximating the function \( T \mapsto \theta(T) \) through a polynomial. In our smart contract, we only consider one threshold with \( \overline{Q} = 5 \). The polynomial approximation of degree 4 is provided in \cref{fig:theta_strasbourg_by_day}.

\begin{figure}[!ht]
\centering
\includegraphics[width=0.6\linewidth]{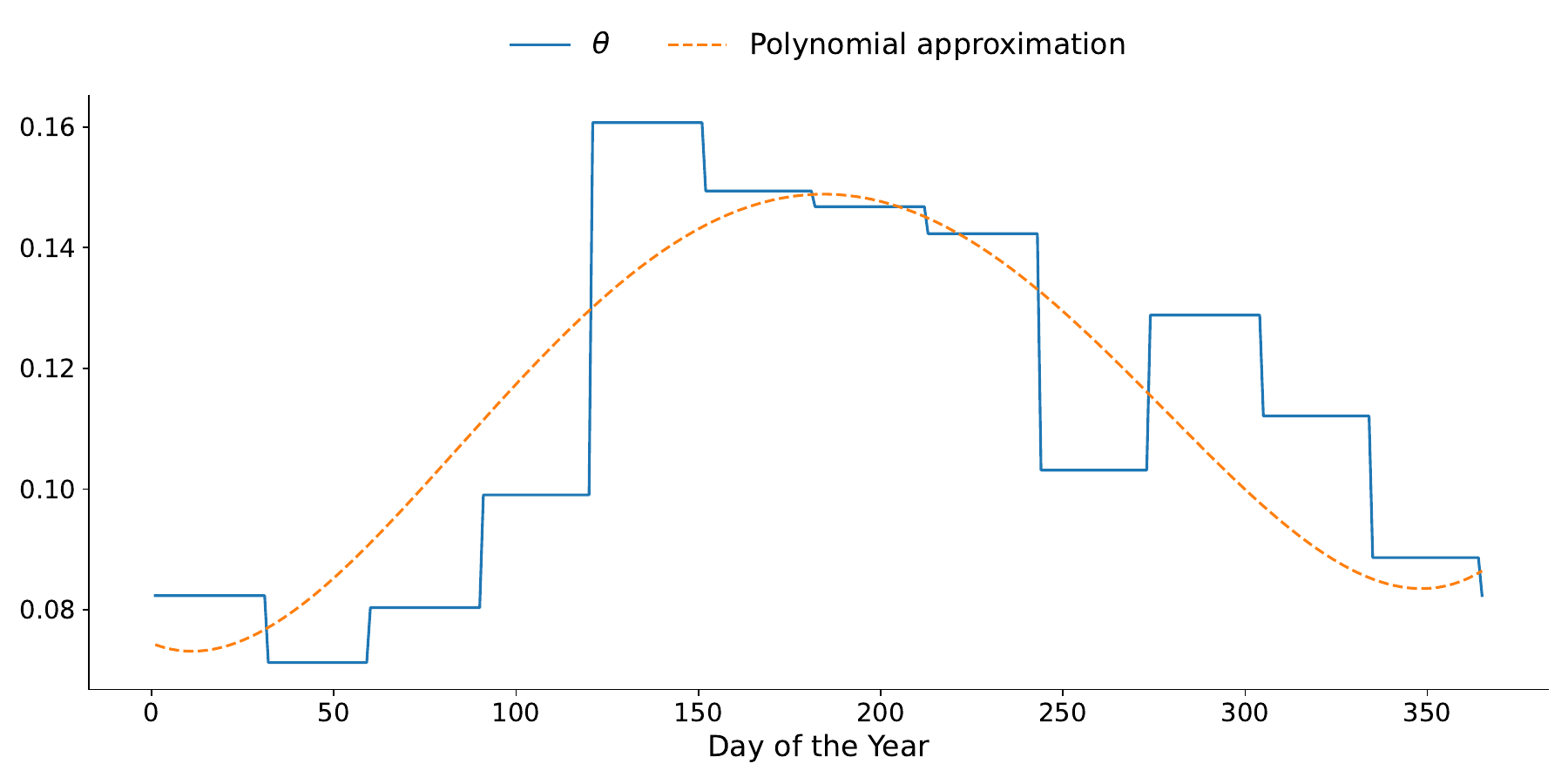}
\caption{Estimation of the probability \( \mathbb{P}(Q_T > 5) \) depending on the day of the year.}
\label{fig:theta_strasbourg_by_day}
\end{figure}

Concretely, we use Python to find the coefficients of the polynomial approximation:

\[
\theta(T) \approx a_0 + a_1 \cdot T + a_2 \cdot T^2 + a_3 \cdot T^3 + a_4 \cdot T^4,
\]

before hardcoding the formula in Solidity. We note in passing that Solidity could allow us to encode a pricing formula that could result from the fitting of a logistic regression model, which is suitable for modeling the probability of occurrence of an event, provided that the model does not include too many covariates. The use of logistic regression for ratemaking purposes of parametric insurance contracts has been explored in \citet{Figueiredo2018}.
\end{remark}

Suppose that \( S \leq S^\ast \), where \( S^\ast \) has been defined in \cref{rem:adverse_selection_linked_to_prediction}. The commercial premium is then given by
\[
\pi_S = (1 + \eta_S) \cdot \theta \cdot l,
\]
using the expectation premium principle, where \( \eta_S > 0 \) is the loading factor at time \( S \) to ensure profitability on average and cover management costs. The safety loading evolves over time according to managerial decisions based on the available information \( \mathcal{F}_t \). It constitutes an \( \mathcal{F}_t \)-measurable stochastic process denoted by \( (\eta_t)_{t \geq 0} \).

\begin{remark}\label{rem:adjust_safety loading for fairness}
An insurance coverage $(S, T, Q, \overline{Q}, l)$ cannot be purchased after time $S^\ast$.
To further mitigate the risk of adverse selection, the safety loading could also be adjusted to incentivize early purchases---specifically, by offering a premium discount to customers who buy the coverage well before $S^\ast$.  
\end{remark}

\subsection{Parametric insurance portfolio over time}\label{ssec:parametric_inusrance_portfolio}
A parametric insurance portfolio is a collection of parametric insurance contracts \(\{(S_i, T_i, Q_i, \overline{Q}_i, l_i)\}_{i \geq 1}\). These contracts may be associated with their own observable quantities \(Q_i := \left(Q_t^{i}\right)_{t \geq 0}\) defined on a common probability space \((\Omega, \mathcal{F}, (\mathcal{F}_t)_{t \geq 0}, \mathbb{P})\).
Let us take a concrete situation in \cref{ex:parametric_insurance_against_rain_4_locations} by following up on \cref{ex:parametric_insurance_against_rain}. 
\begin{ex}\label{ex:parametric_insurance_against_rain_4_locations}
In addition to Strasbourg, we wish to offer coverage against rain episodes in the city of Marseille. We assume that Strasbourg and Marseille are sufficiently far apart (see the map on \cref{fig:france_map}) so that the precipitation levels of these cities are not correlated.
\begin{figure}[!ht]
\centering
\includegraphics[width=0.5\linewidth]{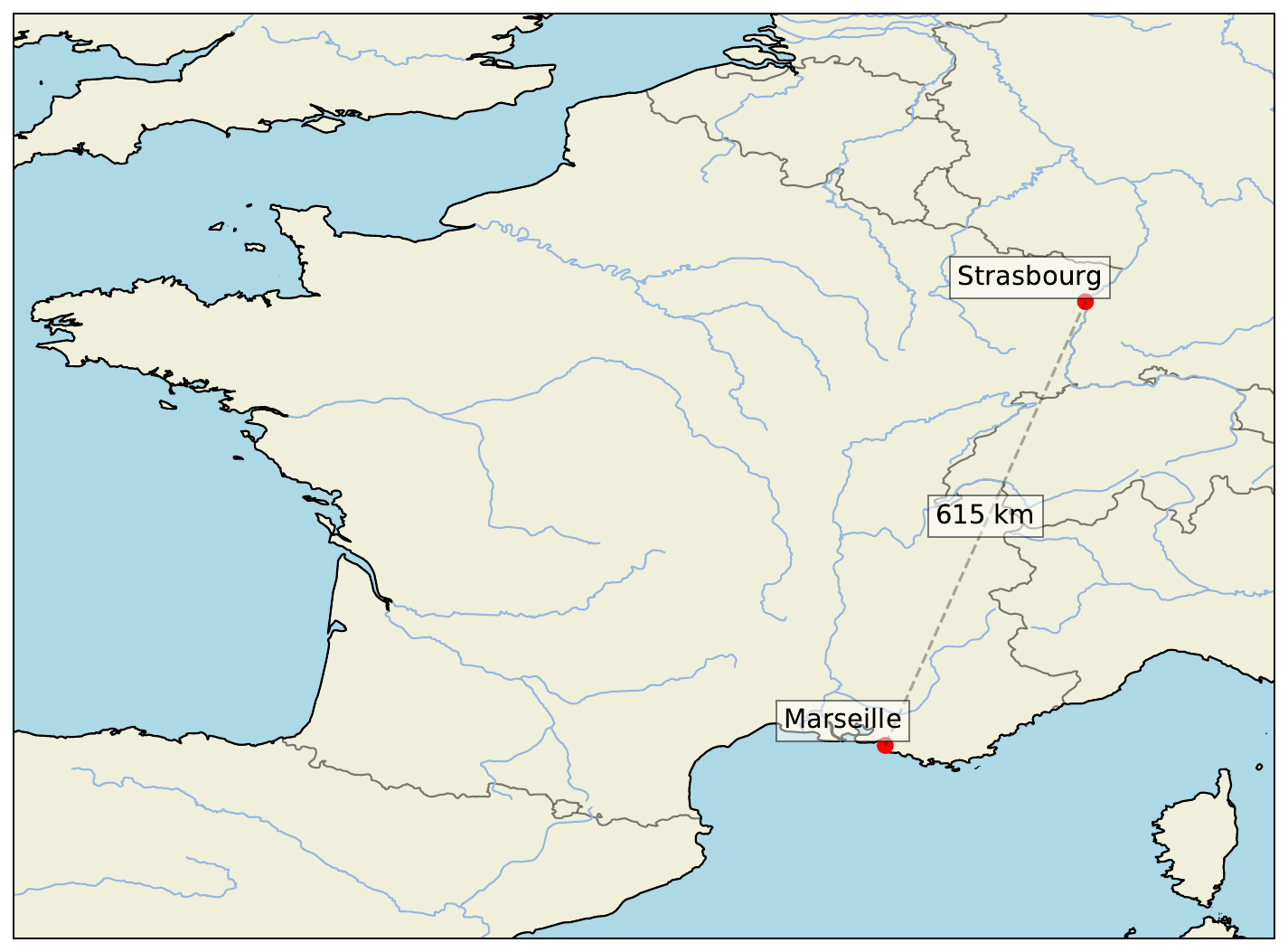}
\caption{Location of Marseille and Strasbourg in France}
\label{fig:france_map}
\end{figure}
Each contract \( i \) is then associated with a location so that \( Q^{i} \in \{Q^{\text{Marseille}}, Q^{\text{Strasbourg}}\} \) and a date \( T_i \) for all \(i\geq 1\). We assume that \( Q^{\text{Marseille}} \) and \( Q^{\text{Strasbourg}} \) are mutually independent. We assume that \( S_i < S_i^\ast \) for each contract so as to model the precipitation height via a compound Poisson-gamma distribution at each location. The models are fitted to the same data as in \cref{ex:parametric_insurance_against_rain} and result in two different risk profiles as illustrated on \cref{fig:theta_by_station_by_month}, where we plot the probability \( \mathbb{P}(Q_T^i > 5) \) as a function of the month of \( T \) for each location $i$.

\begin{figure}[!ht]
\centering
  \includegraphics[width=0.6\linewidth]{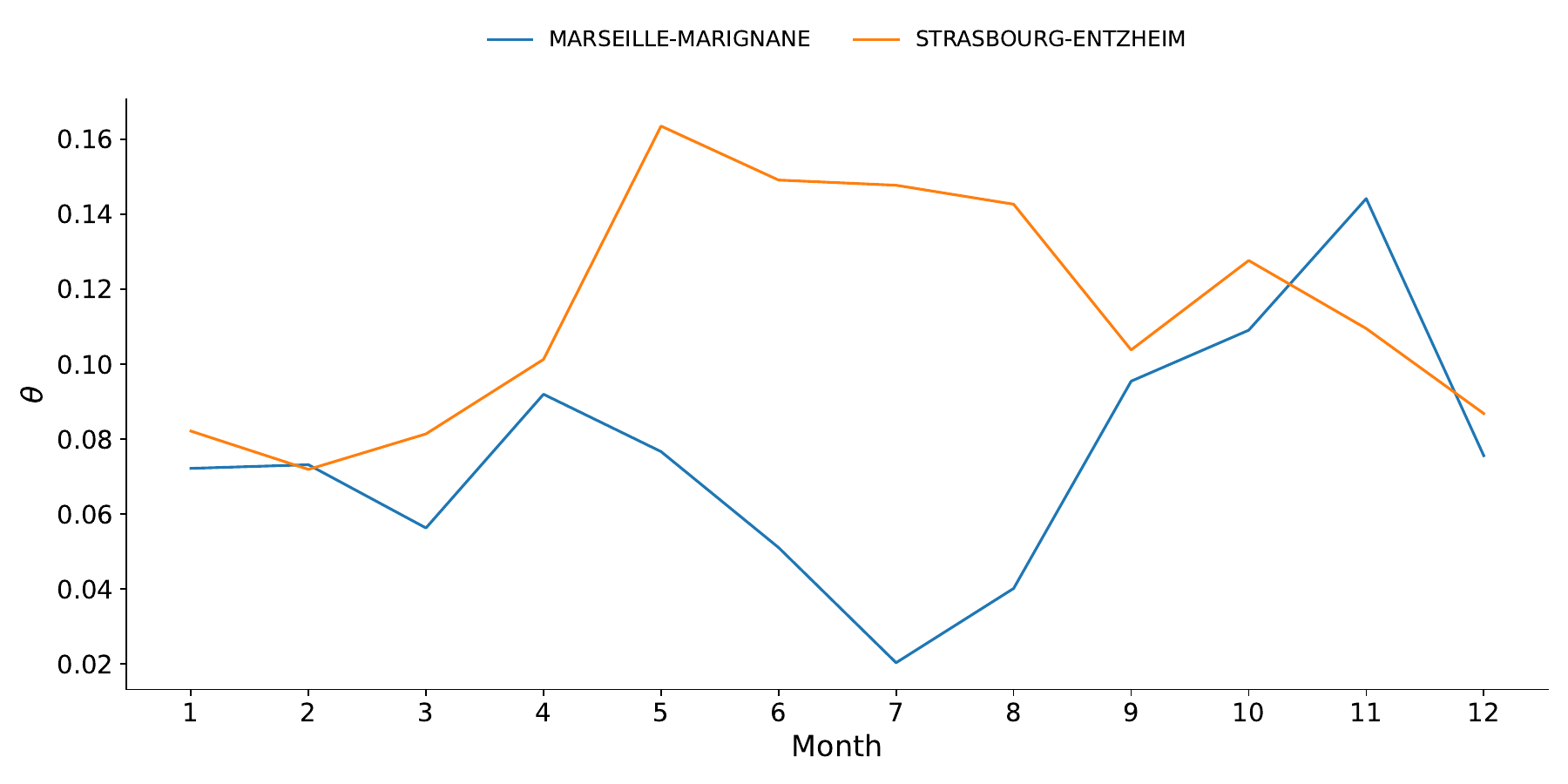}
  \caption{Estimation of the probability $\mathbb{P}(Q_T> 5)$ depending on the month and the location.}
  \label{fig:theta_by_station_by_month}
\end{figure}
\end{ex} 
The number of contracts underwritten up to time \( t \geq 0 \) is modeled as an \(\mathcal{F}_t\)-adapted counting process, defined as
\[
N_0 = 0, \quad N_t = \sum_{i \geq 1} \mathbb{I}_{S_i \leq t}.
\]
The statuses of the contracts are stored in a vector \(\Xi_t = \left(\xi_t^{(1)}, \ldots, \xi_t^{(N_t)}\right)\). Denote by \( Y_i = \mathbb{I}_{Q^i_{T_i} > \bar{Q}_i} \cdot l_i \) the payout of insurance contract \( i \), where \(\left(\mathbb{I}_{Q^i_{T_i} > \bar{Q}_i}\right)_{i = 1, \ldots, N_t}\) is a sequence of Bernoulli variables with parameters 
\[
\theta_i = \mathbb{P}(Q^i_{T_i} > \bar{Q}_i|\mathcal{F}_{S_i})\text{ for } i = 1, \ldots, N_t.
\]

The liability at time \( t > 0 \) is a stochastic process defined as
\begin{equation}\label{eq:liability}
L_0 = 0, \quad L_t = \sum_{i=1}^{N_t} Y_i \mathbb{I}_{S_i \leq t, T_i > t}, \quad \text{for } t \geq 0.
\end{equation}
It corresponds to the future losses given the active policies in our portfolio at time \( t \). 

\begin{remark}\label{rem:bernouilli_approximation}
Note that the assumption of fixed-parameter Bernoulli variables $Y_i$ with parameters is an approximation. In reality, the event probabilities follow stochastic processes, recall the definition \eqref{eq:theta_stochastic_process}. A more precise approach would involve dynamically updating these probabilities when analyzing the liability distribution at time~$t$. However, we adopt this simplification to avoid the prohibitive transaction costs that would arise from real-time updates in a blockchain environment, which would require frequent oracle queries and solvency capital reevaluations.
\end{remark}

The premiums collected to compensate for this liability are given by
\[
\Pi_t = \sum_{i=1}^{N_t} \pi_i \mathbb{I}_{S_i \leq t, T_i > t}, \quad \text{for } t \geq 0,
\]
where
\[
\pi_i = (1+\eta_{S_i}) \cdot \theta_i \cdot l_i, \quad \text{for } i = 1, \ldots, N_t.
\]
We refer to \(\Pi_t\) as the collected but yet to be earned premiums. We define two solvency capitals as
\begin{equation}\label{eq:SCR_MCR}
SCR_t = \text{Quantile}(L_t - \Pi_t; \alpha_{\text{SCR}}), \quad \text{and} \quad
MCR_t = \text{Quantile}(L_t - \Pi_t; \alpha_{\text{MCR}}).
\end{equation}

Here, \( SCR \) stands for Solvency Capital Requirement, and \( MCR \) stands for Minimum Capital Requirement, by analogy with the European directive Solvency II, under which \( \alpha_{\text{SCR}} = 0.995 \) and \( \alpha_{\text{MCR}} = 0.85 \).

\subsection{Solvency Capital Calculation}\label{ssec:solvency_capital_calculation}
\subsubsection{Model points creation to handle dependency}\label{sssec: model_points}
The main challenge associated with calculating the risk capitals \eqref{eq:SCR_MCR} lies in the study of the distribution of the liability \((L_t)_{t \geq 0}\), more specifically the likely dependency among the Bernoulli variables. We show in \cref{ex:dependent_risk} how to gather two correlated contracts into one within the framework of \cref{ex:parametric_insurance_against_rain} and \cref{ex:parametric_insurance_against_rain_4_locations}.
\begin{ex}\label{ex:dependent_risk}
Consider two contracts \(\left(S_{i_1}, T_{i_1}, Q_{i_1}, \overline{Q}_{i_1}, l_{i_1}\right)\) and \(\left(S_{i_2}, T_{i_2}, Q_{i_2}, \overline{Q}_{i_2}, l_{i_2}\right)\) associated with the same location \(Q = Q_{i_1} = Q_{i_2}\) and the same event date \(T = T_{i_1} = T_{i_2}\). The two contracts do not have the same threshold nor the same underwriting time. We assume without loss of generality that \(\overline{Q}_{i_1} \leq \overline{Q}_{i_2}\) and \(S_{i_1} < S_{i_2}\).  We assume that $S_{i_1}$ and $S_{i_2}$ are both sufficiently far from the event date $T$. Denote by

\[
\text{MP}_t = \begin{cases}
\{i_1\}, & \text{if } S_{i_1} \leq t < S_{i_2}, \\
\{i_1, i_2\}, & \text{if } S_{i_2} \leq t < T, \\
\emptyset, & \text{otherwise,}
\end{cases}
\]

the time-dependent set of correlated contracts. The notation "MP" stands for Model Point and corresponds to a group of contracts. These contracts are aggregated through the definition of a payout as

\[
Z_t = \begin{cases}
Y_{i_1}, & \text{if } S_{i_1} \leq t < S_{i_2}, \\
Y_{i_1} + Y_{i_2}, & \text{if } S_{i_2} \leq t < T, \\
0, & \text{otherwise.}
\end{cases}
\]

The probability distribution of \(Z_t\) for \(t \geq S_{i_2}\) is given by

\[
\mathbb{P}(Z_t = 0) = 1 - \theta_{i_1}, \quad \mathbb{P}(Z_t = l_{i_1}) = \theta_{i_1} - \theta_{i_2}, \quad \text{and} \quad \mathbb{P}(Z_t = l_{i_1} + l_{i_2}) = \theta_{i_2}.
\]
\end{ex} 
Following up on \cref{ex:dependent_risk}, for \( t \geq 0 \), we partition our insurance portfolio \(\left\{\left(S_i, T_i, Q_i, \overline{Q}_i, l_i\right)\right\}_{i = 1,\ldots, N_t}\) into \( M_t \) groups \(\text{MP}_{1, t}, \ldots, \text{MP}_{M_t, t}\) of correlated contracts. The classes are associated with independent payouts \( Z_{j,t} \), \( j = 1, \ldots, M_t \). The process \((M_t)_{t \geq 0}\) counts the number of model points defined up to time \( t \geq 0 \). The liability \((L_t)_{t \geq 0}\) is rewritten as
\[
L_t = \sum_{j=1}^{M_t} Z_{j,t} \mathbb{I}_{T_j > t},
\]
where \( T_j \) coincides with the common event date of the contracts in \(\text{MP}_{j,t}\) and
\[
Z_{j,t} = \sum_{i \in \text{MP}_{j,t}} \mathbb{I}_{Q^i_{T_i} > \overline{Q}_i} \cdot l_i.
\]
The liability is now a sum of independent, discrete, and positive random variables. The concept of model points has been introduced in the European directive Solvency II to speed up the calculations of Best Estimate Liabilities associated with life insurance portfolios when using cash-flow projection models. Grouping and aggregating insurance contracts has become a common practice among actuaries, and various methods have been documented in the actuarial science literature, see for instance the works of \citet{Goffard2015,BlanchetScalliet2017,Kiermayer2020,Gweon2020,Gweon2023}. Here, the creation of model points is a workaround for the dependency of the Bernoulli variables, which allows us to study the distribution of the liability in a tractable way as shown in \cref{sssec:solvency_capital_calculation}.

\subsubsection{Recursive calculation of the solvency capitals}\label{sssec:solvency_capital_calculation}
The liability at time \( t > 0 \) is given by
\begin{equation}
L_0 = 0, \quad L_t = \sum_{j=1}^{M_t} Z_{j,t} \mathbb{I}_{T_j > t}, \quad \text{for } t \geq 0,
\end{equation}
where \( T_j \) denotes the common event dates of the contracts in \(\text{MP}_{j, t}\). The number of active model points is denoted by
\[
M_t^{\text{active}} = \sum_{j=1}^{M_t} \mathbb{I}_{T_j > t} \mathbb{I}_{\underset{i \in \text{MP}_{j,t}}{\min} \, S_i \geq t}.
\]
Provided that $M_t^{\text{active}}  = n$, the payouts of \(\text{MP}_1, \ldots, \text{MP}_n\), at time \( t \geq 0 \) are denoted by \( Z_1, \ldots, Z_n \). Consequently, the distribution of the liability \( L_t = \sum_{k=1}^n Z_k \) is represented as the sum of independent random variables. While it is possible to compute the exact distribution using combinatorial analysis, this method becomes impractical as \( n \) increases. We use instead a Fast Fourier Transform (FFT) procedure. After obtaining the probability distribution of \( L_t \), we can compute its cumulative distribution function and inverse it using a bisection algorithm. However, despite its effectiveness, FFT cannot be implemented on the blockchain due to the limitations of smart contract functions, which are unable to encode complex operations. This constraint necessitates performing solvency capital calculations off-chain, requiring intervention from the smart contract manager. Furthermore, any changes in the portfolio at a later time \( s \geq t \) would require recomputing the probability distribution of the liability \( L_s \) to determine the Solvency Capital Requirement (\(\text{SCR}_s\)) or Minimum Capital Requirement (\(\text{MCR}_s\)). 

To address this, we use a normal approximation, justified by the generalized central limit theorem, which is applicable to independent random variables with finite variance. Let \(\mathbb{E}\left(Z_{k}\right) = \mu_{k}\) and \(\mathbb{V}\left(Z_{k}\right) = \sigma_{k}^2\) for \(k = 1, \ldots, n\) be the mean and variance of the random variables \(Z_{k}\). Define \(s_{n}^2 = \sum_{k=1}^n \sigma_{k}^2\) as the sum of these variances.

Since the random variables \((Z_{k})_{k=1,\ldots, n}\) are bounded, they satisfy Lyapunov's condition that reads as 
\[
\lim_{n \to \infty} \frac{1}{s_{n}^{2+\delta}} \sum_{k=1}^n \mathbb{E}\left[|Z_k - \mu_{k}|^{2+\delta}\right] = 0.
\]

Thus, the following convergence result holds: 

\[
\frac{1}{s_{n}} \sum_{k=1}^n (Z_k - \mu_{k}) \overset{\mathcal{D}}{\longrightarrow} \NormalDist(0,1), \quad \text{as } n \to \infty,
\]

where \(\NormalDist(0,1)\) represents the standard normal distribution and \(\overset{\mathcal{D}}{\longrightarrow}\) denotes convergence in distribution. We note that
\[
L_t - \Pi_t = \sum_{k=1}^n \left(Z_k - \sum_{i \in \text{MP}_k} (1 + \eta_{S_i}) \cdot \theta_i \cdot l_i \right) = s_n \left( \frac{1}{s_n} \sum_{i=1}^n (Z_i - \mu_i) \right) - \sum_{k=1}^n \sum_{i \in \text{MP}_k} \eta_{S_i} \cdot \theta_i \cdot l_i.
\]

We deduce the following approximations for the SCR and MCR:

\begin{equation}\label{eq:normal_approximation_SCR_MCR}
\text{SCR}_t \approx s_n \cdot q_{\alpha_{\text{SCR}}} - \sum_{k=1}^n \sum_{i \in \text{MP}_k} \eta_{S_i} \cdot \theta_i \cdot l_i, \quad \text{and} \quad \text{MCR}_t \approx s_n \cdot q_{\alpha_{\text{MCR}}} - \sum_{k=1}^n \sum_{i \in \text{MP}_k} \eta_{S_i} \cdot \theta_i \cdot l_i,
\end{equation}

where \( q_\alpha \) is the quantile of order \( \alpha \) of the standard normal distribution. The accuracy of the approximation in \eqref{eq:normal_approximation_SCR_MCR} depends on the convergence of the Central Limit Theorem (CLT) as \( n \) grows large. A common way to improve this approximation is through the Cornish-Fisher expansion, as seen in \citet{Cornish1938,Fisher1960}. We limit ourselves to the third and fourth-order expansions, which involve replacing the quantile \( q_\alpha \) by:

\begin{equation}\label{eq:CF_order_3_and_4}
q_\alpha + \gamma_1 \frac{q_\alpha^2 - 1}{6}, \quad \text{and} \quad q_\alpha + \gamma_1 \frac{q_\alpha^2 - 1}{6} + \gamma_2 \frac{q_\alpha^3 - 3q_\alpha}{24} - \gamma_1^2 \frac{2q_\alpha^3 - 5q_\alpha}{36},
\end{equation}

respectively, where \( \gamma_1 \) and \( \gamma_2 \) denote the skewness and excess kurtosis of \( L_t \). Note that the approximation in \eqref{eq:normal_approximation_SCR_MCR} is referred to as the second-order Cornish-Fisher approximation. We then decide on a threshold \( \bar{n} \) such that if \( n > \bar{n} \), we compute the solvency capital using the approximations in \eqref{eq:normal_approximation_SCR_MCR} or \eqref{eq:CF_order_3_and_4}. Otherwise, we fall back to:

\[
\text{SCR}_t = \text{MCR}_t = \sum_{k=1}^{N_t} l_k.
\]

The determination of such a threshold can be based on a brief simulation study, as illustrated in \cref{ex:simulation-cornish}.

\begin{ex}\label{ex:simulation-cornish}
We sample a synthetic parametric insurance portfolio \(\{(S_i, T_i, Q_i, \overline{Q}_i, l_i)\}_{i \geq 1}\) in order to yield exactly \( n = 100 \) model points. To achieve this, we select $100$ combinations of location and date. For each combination, we draw uniformly at random from \(\{1, 2, 3, \ldots, 10\}\) a number of parametric insurance contracts to be aggregated in a model point.
The threshold is fixed to  \(\overline{Q}_i = 5\) and the compensations are drawn at random as:

\[
l_i \sim \text{Unif}(\{5, 10, 15, 20\}), \quad \text{for } i \geq 1.
\]

The first five rows of such a synthetic portfolio are provided in \cref{tab:fake_parametric_insurance_portfolio}.

\begin{table}[!ht]
\scriptsize
\centering
\begin{tabular}{llrrl}
\toprule
 \textbf{S}          & \textbf{T}          &   \textbf{Q} &   \textbf{l} & \textbf{station}             \\
 2024-11-23 & 2025-01-08 &   5 &   5 & MARSEILLE-MARIGNANE \\
 2024-12-23 & 2025-01-08 &   5 &   5 & STRASBOURG-ENTZHEIM \\
 2024-07-06 & 2025-01-15 &   5 &   5 & STRASBOURG-ENTZHEIM \\
 2024-01-26 & 2025-01-10 &   5 &  20 & STRASBOURG-ENTZHEIM \\
 2024-01-06 & 2025-01-09 &   5 &  10 & MARSEILLE-MARIGNANE \\
\bottomrule
\end{tabular}
\caption{Characteristics of the first five parametric insurance contracts}
\label{tab:fake_parametric_insurance_portfolio}
\end{table}
We simulate 100 synthetic portfolios and calculate the relative errors between the exact and approximated solvency capitals for an increasing number of model points, \( n \). The order of the quantiles are set to \( q_{\alpha_{\text{MCR}}} = 0.85 \) and \( q_{\alpha_{\text{SCR}}} = 0.995 \). The results of these simulations are displayed in \cref{fig:relative_error_SCR_MCR}.

\begin{figure}[!ht]
  \begin{center}
    \subfloat[Relative error on MCR]{
      \includegraphics[width=0.5\textwidth]{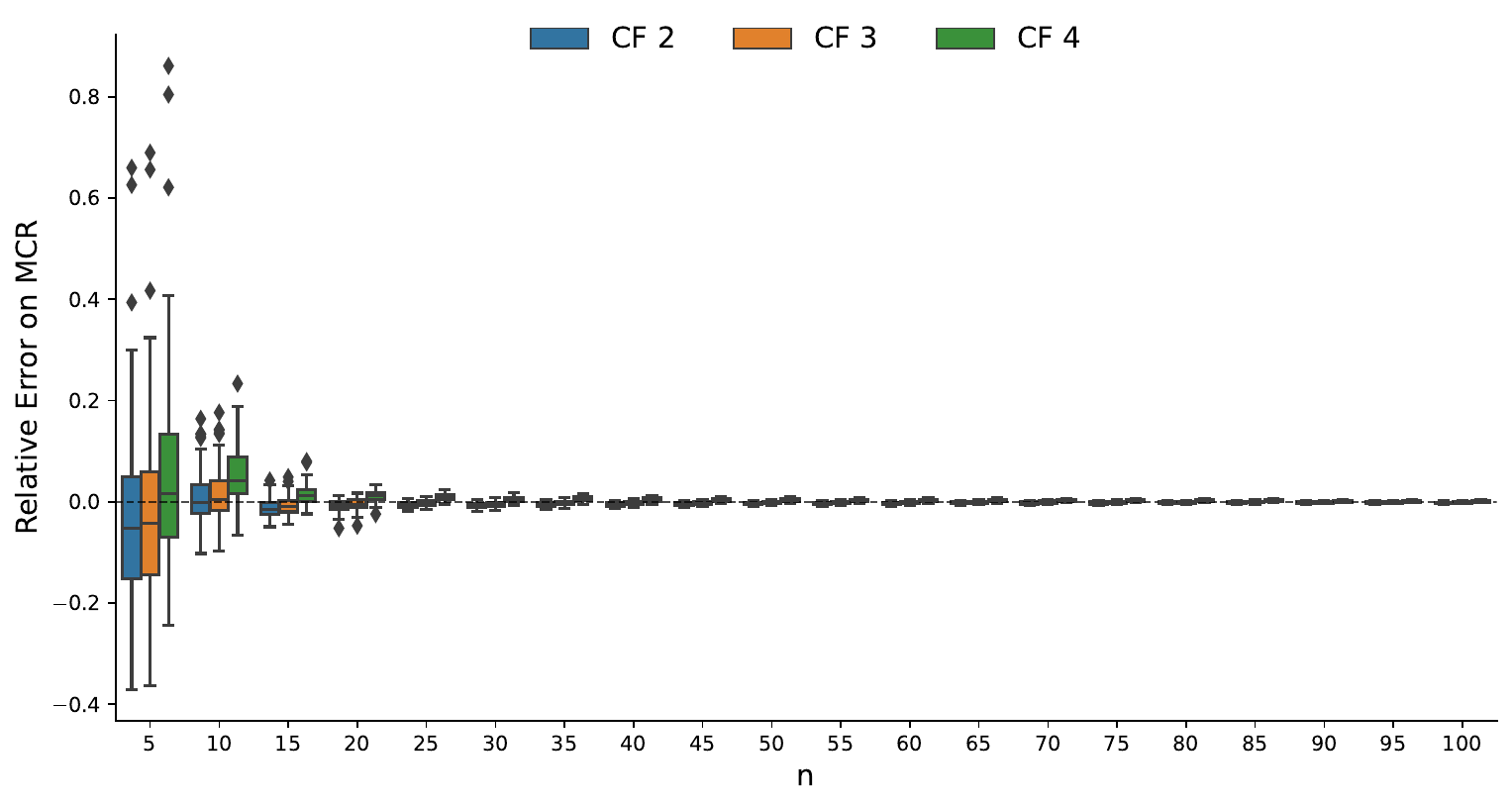}
      \label{sub:MCR}
    }
    \subfloat[Relative error on SCR]{
      \includegraphics[width=0.5\textwidth]{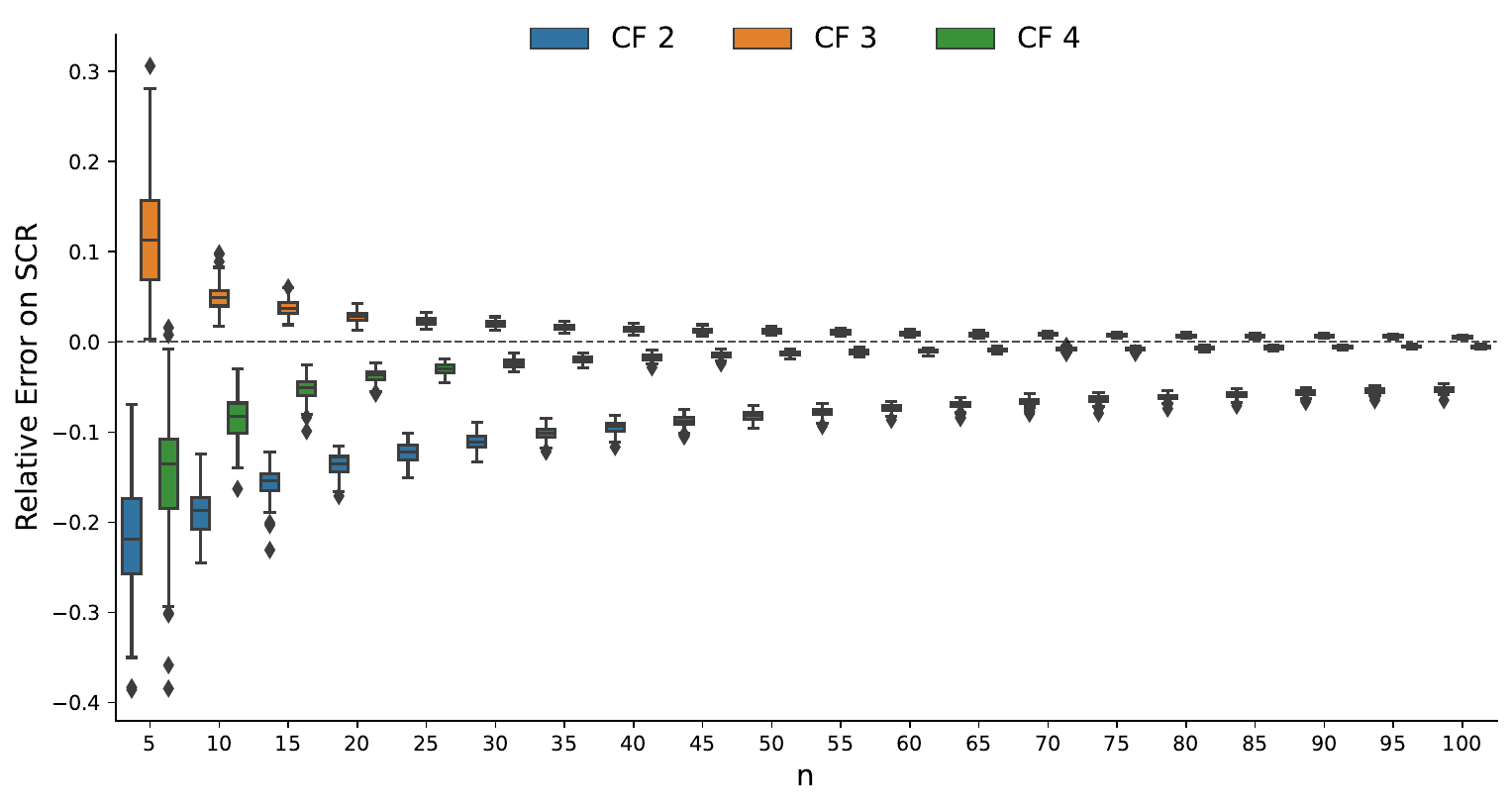}
      \label{sub:SCR}
    }
    \caption{Relative error on MCR and SCR depending on the number of models points in the porttfolio.}
    \label{fig:relative_error_SCR_MCR}
  \end{center}
\end{figure}

The normal approximation becomes acceptable for the MCR relatively quickly, but the relative error for the SCR converges more slowly. Overall, adjusting for skewness is beneficial for estimating both the MCR and SCR. The level of error falls below 5\% when \( n \geq 15 \) and drops below 1\% for \( n \geq 30 \). If we consider a 5\% error level to be acceptable, then the third-order Cornish-Fisher approximation could be used as soon as the number of model points reaches 15.
\end{ex}
\begin{remark}\label{rem:limiting_correlated_risks}
It is common practice among insurers to limit the underwriting of highly correlated risks. This can be easily implemented by preventing underwriting if the contract is aggregated within a model point that already contains too many contracts. Note that we implicitly do this in \cref{ex:simulation-cornish}, as the number of correlated risks is limited to 10. Additionally, the condition could be based on the maximum possible loss associated with the model point.
\end{remark}

The conclusions and resulting recommendations of \cref{ex:simulation-cornish} are valid for the specific case under consideration. A similar study could be conducted to adapt the threshold for the number of model points and guide the selection of an appropriate order for the Cornish-Fisher approximation for another situation.

\section{Smart contract description}\label{sec:smart_conract_for_parametric_insurance}

A smart contract is a system whose state evolves according to the actions of blockchain users. The roles of the users of our smart contract are outlined in \cref{ssec:roles}. The stochastic processes used to monitor the state of the system are defined in \cref{ssec:state_variable}. We introduce the methods that allow users to interact with the smart contract in \cref{ssec:smart_contract_methods}.

\subsection{Role distribution}\label{ssec:roles}
The participants in our blockchain insurance scheme include the smart contract owner, the investors (also referred to as surplus providers), and the policyholders. The processes of underwriting and claim assessment are fully automated within the smart contract protocol. The role of our investors is analogous to that of the reinsurers in \citet{Cousaert2022}.

\subsubsection{Smart contract owner}\label{sssec:contract_owner}
The smart contract owner is responsible for designing the insurance protocol by writing the Solidity code for the smart contract. While deploying the smart contract on the Ethereum blockchain, the owner initializes three parameters: the premium loading $\eta$, the risk levels $\alpha_{\text{SCR}}$ and $\alpha_{\text{MCR}}$ associated with the solvency capitals, as defined in \cref{sec:parametric_insurance}, and the lower bound $m$ for the number of active model points to use the normal approximation of \cref{ssec:solvency_capital_calculation} to calculate the solvency capitals. Additionally, the contract owner acts as the claim assessor or data oracle by inputting the observed quantity at the time of the event—in this case, the precipitation height. In the blockchain ecosystem, an oracle refers to an off-chain data provider that a smart contract can query. This feature is not implemented in our current solution, it would be beneficial to do so in order to further automate our workflow.

\subsubsection{Surplus providers}\label{sssec:RP}
Surplus providers, also known as investors, contribute financially to the smart contract to constitute the solvency capital. These contributions are denominated in \text{ETH}, the native cryptocurrency of the Ethereum blockchain. In return for their investment, investors receive tokens that represent their proportional share of the surplus. Surplus providers have the option to withdraw a portion of their funds by burning these tokens in exchange for an equivalent amount of \text{ETH}, with the exchange rate defined in the \cref{ssec:state_variable}. Withdrawals are permitted provided that the funds remaining in the smart contract do not fall below a specified threshold, known as the Solvency Capital Requirement (SCR).

\subsubsection{Policyholders}\label{sssec:policyholders}
Policyholders seek financial protection against adverse events through our smart contract. Upon the occurrence of such events, the smart contract pays a lump-sum compensation in \text{ETH}. To obtain this coverage, policyholders must pay a premium, also in \text{ETH}, to the smart contract.

An insurance agreement is only concluded if the SCR, after accounting for the new risk, does not exceed the funds currently locked in the smart contract. Should the settlement of a claim lead to the smart contract's funds falling below the MCR, the system enters a state of bankruptcy. In such instances, priority is given to reimbursing policyholders with the remaining funds. Any funds that remain after all policyholder reimbursements are distributed to investors, proportional to their token holdings.


\subsection{State variables}\label{ssec:state_variable}
The state variables of the smart contract include the stochastic processes defined in \cref{sec:parametric_insurance}, supplemented by additional processes defined on the filtered probability space \((\Omega, \mathcal{F}, (\mathcal{F})_{t \geq 0}, \mathbb{P})\).

Upon deployment on the Ethereum blockchain, a smart contract is assigned an Ethereum address, enabling it to send and receive \text{ETH}. The balance \((B_t)_{t \geq 0}\) increases when investors contribute funds or when premiums are paid for underwritten policies. Conversely, the balance decreases upon investor withdrawals or compensation payments to policyholders. Let \((\tau_k^+)_{k \geq 1}\) and \((\tau_k^-)_{k \geq 1}\) denote the times of investor deposits and withdrawals, respectively, with associated amounts \((x_k^+)_{k \geq 1}\) for deposits and \((x_k^-)_{k \geq 1}\) for withdrawals. The balance can be expressed as:

\[
B_0 = 0, \quad \text{and} \quad B_t = \sum_{k \geq 1} \mathbb{I}_{\tau_k^+ \leq t} x_k^+ - \sum_{k\geq 1} \mathbb{I}_{\tau_k^- \leq t} x_k^- + \sum_{i \geq 1} \mathbb{I}_{S_i \leq t} \pi_i - \sum_{i \geq 1} \mathbb{I}_{T_i \leq t} \mathbb{I}_{Q_{T_i}> \bar{Q}_i} l_i, \quad \text{for } t \geq 0.
\]

We define \((X_t)_{t \geq 0}\) as the surplus of the smart contract, representing the funds available to underwrite new insurance policies. Unlike the balance, the surplus does not increase immediately upon underwriting a new contract. Instead, the premium is added at the policy's resolution, when it is considered "earned." Thus, the surplus is given by:

\[
X_0 = 0, \quad \text{and} \quad X_t = \sum_{k \geq 1} \mathbb{I}_{\tau_k^+ \leq t} x_k^+ - \sum_{k \geq 1} \mathbb{I}_{\tau_k^- \leq t} x_k^- + \sum_{i \geq 1} \mathbb{I}_{T_i \leq t} \pi_i - \sum_{i \geq 1} \mathbb{I}_{T_i \leq t} \mathbb{I}_{Q_{T_i}> \bar{Q}_i} l_i, \quad \text{for } t \geq 0.
\]

Note that \(B_t \geq X_t\) for every \(t \geq 0\). The surplus \((X_t)_{t \geq 0}\) is compared against \(\text{SCR}_t\) and \(\text{MCR}_t\). If \(X_t \leq \text{SCR}_t\), no new policies may be underwritten, and investors are prohibited from withdrawing funds. If \(X_t \leq \text{MCR}_t\), the smart contract ceases operations, and the remaining balance is redistributed—first to policyholders and then to investors, as detailed in \cref{ssec:settlement}. Active insurance policies at the time of termination are cancelled.

Let \((A_t)_{t \geq 0}\) denote the total number of distinct Ethereum addresses that have transferred funds to the smart contract and received protocol tokens. Let

\[
Y_t = \begin{pmatrix} Y_t^{(1)} & \ldots & Y_t^{(A_t)} \end{pmatrix},
\]

be a vector representing the token holdings of the smart contract participants. The total supply of tokens is given by \(\overline{Y}_t = \sum_{i = 1}^{A_t} Y_t^{(i)}\). The exchange rate of tokens against \text{ETH} is modeled by the stochastic process \((r_t)_{t \geq 0}\), defined as:

\[
r_0 = 1, \quad r_t = \frac{X_t}{\overline{Y}_t}, \quad \text{for } t \geq 0.
\]

The exchange rate must be updated whenever \(X_t\) or \(\overline{Y}_t\) changes. Lastly, denote by 
\[
\Lambda_0 = 0, \quad \Lambda_t = \sum_{i=1}^{N_t} l_i \mathbb{I}_{S_i \leq t, T_i > t}, \quad \text{for } t \geq 0,    
\]
the sum of all the potential compensations to be paid.

  
\subsection{Method of the smart contract}\label{ssec:smart_contract_methods}
The state of a smart contract changes according to the actions of blockchain users. In our parametric insurance example, these users include the contract owner, surplus providers, and policyholders (see \cref{ssec:roles}). Participants interact with the smart contract through specific functions, known as methods, which modify the system's state. The remainder of this section describes each possible action and its impact on the stochastic processes introduced thus far.
\subsubsection{Funding the contract}\label{ssec:fund}
Assume that an investor transfers \( x \) \text{ETH} to the contract at time \( t > 0 \). As a result, both the balance and the surplus of the contract increase as follows:

\[
B_{t + h} = B_t + x, \quad X_{t + h} = X_t + x,
\]

where \( h > 0 \) represents an infinitesimal time lapse.

If the investor is new, the count of distinct addresses increases, such that \( A_{t+h} = A_t + 1 \); otherwise, \( A_{t+h} = A_t \). The \( j \)-th component of the vector \( (Y_t)_{t \geq 0} \), where \( j \in \{1, \ldots, A_{t+h}\} \), is updated as:

\[
Y_{t+h}^{(j)} = Y_t^{(j)} + y,
\]

where \( y = x / r_t \).

\subsubsection{Underwriting a parametric insurance policy}
A customer wishes to purchase an insurance contract \((S, T, Q, \overline{Q}, l)\) at time \(t = S\) such that \(S < S^\ast\). The premium for such a parametric insurance contract is given by:

\[
\pi = (1 + \eta_t) \cdot l \cdot \theta,
\]

where \(\theta = \mathbb{P}(Q_T > \overline{Q} | \mathcal{F}_S)\). The contract can only be underwritten if the surplus of the smart contract is sufficient to cover the resulting increase in the SCR.

The liability, the cumulative premium collected, and the number of active model points would then be updated to:

\[
\tilde{\Lambda}_{t} = \Lambda_t + l , \quad \tilde{L}_{t} = L_t + l \cdot \mathbb{I}_{Q_T > \bar{Q}}, \quad  \tilde{\Pi}_{t + h} = \Pi_t + \pi,
\] 
and 
\[\tilde{M}^{\text{active}}_{t} = \begin{cases}
M^{\text{active}}_{t},& \text{if  the new contract is grouped into an active model point}, \\
M^{\text{active}}_{t} + 1,&\text{otherwise}. 
\end{cases}
\]
We calculate the updated solvency capital requirement:
\[
\widetilde{\text{SCR}}_t = \begin{cases}
\text{Quantile}(\tilde{L}_t - \tilde{\Pi}_t; \alpha_{\text{SCR}}),& \text{ if  } \tilde{M}^{\text{active}}_{t} \geq m,\\
\Lambda_t,&\text{otherwise.}
\end{cases}
\]

and underwrite the contract only if:

\begin{equation}\label{eq:underwriting_condition}
X_t \geq \widetilde{\text{SCR}}_t.
\end{equation}

If condition \eqref{eq:underwriting_condition} holds, the number of contracts, liability, premium collected, and number of active model points are updated as follows:

\[
N_{t+h} = N_t + 1, \quad \Lambda_{t+h} = \Lambda_t + l,\quad L_{t+h} = L_t + l \cdot \mathbb{I}_{Q_T > \bar{Q}}, \quad \Pi_{t+h} = \Pi_t + (1 + \eta_t) \cdot l \cdot \theta \text{, and  }M_{t+h}^{\text{active}} = \tilde{M}^{\text{active}}_{t}
\]

The SCR and MCR are also updated as:

\[
\text{SCR}_{t+h} = \begin{cases}
\text{Quantile}(L_{t+h} - \Pi_{t+h}; \alpha_{\text{SCR}}),& \text{ if  } M^{\text{active}}_{t+h} \geq m,\\
\Lambda_t,&\text{otherwise,}
\end{cases}
\]
and 
\[
\text{MCR}_{t+h} = \begin{cases}
\text{Quantile}(L_{t+h} - \Pi_{t+h}; \alpha_{\text{MCR}}),& \text{ if  } M^{\text{active}}_{t+h} \geq m,\\
\Lambda_t,&\text{otherwise.}
\end{cases}
\]
We add a component:

\[
\xi_{t+h}^{(N_{t+h})} = 0,
\]

to the vector \(\Xi_t\) that gathers the statuses of the insurance policies. The balance of the contract is updated as:

\[
B_{t+h} = B_t + (1 + \eta_t) \cdot \theta \cdot l.
\]

\subsubsection{Withdrawing from the smart contract and burning tokens}\label{ssec:withdraw}
An investor \( j \in \{1, \ldots, A_t\} \) may withdraw some of her funds by burning protocol tokens at time \( t > 0 \). This action is subject to the conditions that the participant holds enough protocol tokens \( Y_t^{(j)} \), the smart contract has a sufficient balance \( B_t \), and the surplus \( X_t \) remains adequate.

Assume that participant \( j \) wishes to burn \( y \) tokens at time \( t > 0 \). Upon burning, the participant should receive \( x = y \cdot r_t \) \text{ETH}. The transaction is allowed if the following conditions are met:

\begin{equation}\label{eq:withdrawal_conditions}
y \leq Y_t^{(j)}, \quad x \leq B_t, \quad \text{and} \quad x < X_t - \text{SCR}_t.
\end{equation}

Provided that the conditions in \eqref{eq:withdrawal_conditions} are satisfied, we update the contract balance and the surplus as follows:

\[
B_{t + h} = B_t - x, \quad \text{and} \quad X_{t + h} = X_t - x.
\]

We further update the token balance of investor \( j \) with:

\[
Y_{t+h}^{(j)} = Y_t^{(j)} - y.
\]

\begin{remark}
Since the events that trigger compensation are predictable in the short term, it is essential to impose timing constraints on investors' fund withdrawals. One practical solution would be to require investors to submit withdrawal orders, which the smart contract owner would validate before processing. Under this approach, investors would indicate their intention to withdraw funds at a future date and specify the desired amount in advance. Note that this feature has not been implemented in our current project.

\end{remark}

\subsubsection{Contract resolution and claim settlement}\label{ssec:settlement}
At time \( t = T_i \), the contract \((S_i, T_i, Q_i, \overline{Q}_i, l_i)\), for some \( i = 1, \ldots, N_t \), is settled as it becomes known whether the event \( Q_{T_i} > \overline{Q}_i \) has occurred. The liabilities and the premum collected are updated as follows:
\[
\tilde{\Lambda}_t = \Lambda_t - l_i\text{, }\tilde{L}_t = L_t - l_i, \quad \tilde{\Pi}_{t} = \Pi_t - (1 + \eta_{S_i}) \cdot l_i \cdot \theta_i.
\]
The balance of the contract remains unchanged, \( \tilde{B}_t = B_t \), while the surplus is updated as:

\[
\tilde{X}_{t} = \begin{cases}
X_t + (1 + \eta_{S_i}) \cdot \theta_i \cdot l_i,&\text{ if  }Q_{T_i} \leq  \overline{Q}_i,\\
X_t + (1 + \eta_{S_i}) \cdot \theta_i \cdot l_i - l_i,&\text{ otherwise}.
\end{cases}
\]
Recall that \( S_i \) is the time at which contract \( i \) was underwritten. The number of active model points is updated as follows:

\[\tilde{M}^{\text{active}}_{t} = \begin{cases}
M^{\text{active}}_{t} - 1,& \text{if the contract was the last contract of a given model point}, \\
M^{\text{active}}_{t},&\text{otherwise}. 
\end{cases}
\]
The SCR and MCR are also updated as:

\[
\tilde{\text{SCR}}_{t} = \begin{cases}
\text{Quantile}(\tilde{L}_{t} - \tilde{\Pi}_{t}; \alpha_{\text{SCR}}),& \text{ if  } \tilde{M}^{\text{active}}_{t} \geq m,\\
\tilde{\Lambda}_t,&\text{otherwise,}
\end{cases}
\]
and 
\[
\tilde{\text{MCR}}_{t} = \begin{cases}
\text{Quantile}(\tilde{L}_{t} - \tilde{\Pi}_{t}; \alpha_{\text{MCR}}),& \text{ if  } \tilde{M}^{\text{active}}_{t} \geq m,\\
\tilde{\Lambda}_t,&\text{otherwise.}
\end{cases}
\]
At this stage we distinguish two cases. First assume that $\tilde{X}_{t} > \tilde{\text{MCR}}_{t}$ then all the processes are updated for good with 
    \[
        B_{t+h} = \tilde{B}_t\text{, }X_{t+x} = \tilde{X}_t\text{, }\Lambda_{t+h} = \tilde{\Lambda}_t\text{, }L_{t+h} = \tilde{L}_t\text{, }\Pi_{t+x} = \tilde{\Pi}_t\text{, }\text{SCR}_{t+h} = \tilde{SCR}_t\text{, }\text{MCR}_{t+x} = \tilde{\text{MCR}}_t\text{, and }M_{t+h}^{active} = \tilde{M}^{\text{active}}_{t}.
    \]
The premium for contract \( i \) is now considered "earned." The exchange rate is updated as:

\[
r_{t+h} = \frac{X_{t+h}}{\overline{Y}_{t+h}},
\]
where it is noted that the total supply of protocol tokens remains unchanged between \( t \) and \( t+h \).

Now suppose that the surplus downcrosses the MCR as \(\tilde{X}_{t} \leq \tilde{\text{MCR}}_{t}\). If this occurs, the smart contract is reset. The outstanding balance is first used to reimburse the policyholders of active contracts whose status become "cancelled". The total amount is therefore given by:

\[
x = \sum_{i = 1}^{N_{t}} \mathbb{I}_{\xi^{(i)}_t = 0} \left(1 + \eta_{S_i}\right) \cdot \theta_i \cdot l_i.
\]

If \( \tilde{X}_{t} > x \), the premiums are fully refunded, and the remaining wealth, \( \tilde{X}_{t} - x \), is distributed to the investors according to their token holdings, using the updated exchange rate:

\[
\tilde{r}_{t} = \frac{\tilde{X}_{t} - x}{\bar{Y}_t}.
\]

The smart contract transfers:

\[
x_j = \tilde{r}_{t} \cdot Y^{(j)}_t,
\]

to all investors \( j = 1, \ldots, A_{t} \).

If \( \tilde{X}_{t} \leq x \), the policyholders are reimbursed pro rata based on the premiums they paid. Specifically, we define weights for the policyholders with active contracts as:

\[
w_i = \frac{\mathbb{I}_{\xi^{(i)}_t = 0} \left(1 + \eta_{S_i}\right) \cdot \theta_i \cdot l_i}{\sum_{i = 1}^{N_{t}} \mathbb{I}_{\xi^{(i)}_t = 0} \left(1 + \eta_{S_i}\right) \cdot \theta_i \cdot l_i},
\]

and transfer:

\[
x_i = w_i \cdot \tilde{X}_{t},
\]

to the policyholder associated to policy \( i \) for \( i = 1, \ldots, N_{t} \).

After completing these transfers, the smart contract variables are reset to their initial state with:
\[
B_{t+h} = 0, \quad X_{t+h} = 0,\quad \Lambda_{t+h} = 0, \quad L_{t+h} = 0, \quad \Pi_{t+h} = 0, \quad \text{SCR}_{t+h} = 0, \quad \text{MCR}_{t+h} = 0, \quad r_{t+h} = 1\text{, and }M_{t+h}^{active} = 0.
\]
The number of contracts and the number of token holders remain unchanged:
\[
N_{t+h} = N_t, \quad A_{t+h} = A_t.
\]
However, all token holdings are reset to zero:
\[
Y^{(j)}_{t+h} = 0, \quad \text{for all } j = 1, \ldots, A_{t+h}.
\]

The practical implementation of the smart contract using the programming language Solidity is documented in the online accompaniment of this paper\footnote{\url{https://github.com/LaGauffre/smart_parametric_insurance/blob/main/latex/sup_material_blockchain_parametric_insurance.pdf}}.

\section{Interaction with the smart contract}\label{sec:smart_contract_interaction}
During the development phase, a smart contract is not immediately deployed on the Ethereum mainnet. Instead, developers use testnets to simulate the behavior of the real blockchain, ensuring that the smart contract operates as intended. For our project, we deployed the smart contract on the Ethereum testnet known as Sepolia. Sepolia includes a block explorer called Etherscan\footnote{See \url{https://sepolia.etherscan.io/}}, which provides an API for data retrieval. Our contract, named \texttt{InsuranceLogic}, is located at the address:
\[
\texttt{0xbe035cf1367c45A0C9517969F5ABDd3abF743ae7}.
\]
It can be viewed on Etherscan\footnote{see \url{https://sepolia.etherscan.io/address/0xbe035cf1367c45A0C9517969F5ABDd3abF743ae7}}. The complexity of the insurance mechanism necessitated the deployment of two additional smart contracts. \texttt{PricingLogic} located at the address
\[
\texttt{0xb5a208E3d8c74464d7a70C9B7219880097c6DE81},
\]
calculates the premium and \texttt{ModelPointsLogic} located at
\[
\texttt{0xA1Ad49F8a7e8781A0488029eF7AC471901131Fce},
\]
computes the solvency capitals. The Solidity code for these smart contracts is thoroughly discussed in the online supplementary materials that accompany this paper.\footnote{\url{https://github.com/LaGauffre/smart_parametric_insurance/}}. 

The deployment of the contract was carried out using Remix IDE\footnote{See \url{https://remix.ethereum.org/}} which is  an open-source, web-based integrated development environment specifically designed for writing, testing, and deploying smart contracts on the Ethereum blockchain. We first compile the Solidity code before deploying it to the blockchain. Once deployed, the contract is associated with an Ethereum address, enabling interaction through Remix IDE by calling functions and checking the values of the state variables.

Each action on the testnet incurs a cost denominated in Sepolia testnet tokens, which act as a substitute for ETH. Payments are made through wallets. We use MetaMask\footnote{See \url{https://metamask.io/}}, a popular cryptocurrency wallet and gateway to blockchain applications. MetaMask serves as a browser extension or mobile app that enables users to interact with Ethereum-based decentralized applications (dApps) directly from their web browser or smartphone.

To test the smart contract, three Ethereum accounts were created to represent the roles of the smart contract owner, surplus provider, and policyholder. The addresses of these accounts (plus that of the smart contract) are provided in \cref{tab:eth_address}.

\begin{table}[ht]
\footnotesize
\begin{center}
\begin{tabular}{lcr}
\toprule
ETH Address &\phantom{abc}& Role  \\ 
\texttt{0xE8e79B8B8c0481fa33a8E0fcA902ad5754BfE1C3}&& \textit{owner} \\ 
\texttt{0x2CF8ed1664616483c12Ef3113f6F62E68f1a810A}&& \textit{surplus provider} \\ 
\texttt{0xd34a37613A382bA503f1599F514C9788dF3659C4}&& \textit{policyholder}   \\ 
\texttt{0xbe035cf1367c45A0C9517969F5ABDd3abF743ae7}&& \textit{smart contract} \\ 
\bottomrule
\end{tabular}%
\end{center}
\caption{Ethereum addresses of the smart contract users.}
\label{tab:eth_address}
\end{table}

Our smart contract allows for the underwriting of parametric insurance contracts to compensate policyholders if the height of precipitation exceeds a threshold \(\bar{Q} = 5\) on a particular day in the year 2025 in either Marseille or Strasbourg. The Cornish-Fisher approximation is employed as soon as we reach 5 model points. The day of the year is represented as an integer between 1 and 365. We designed the following scenario to illustrate the functions available in the smart contract described in \cref{sec:smart_conract_for_parametric_insurance}.

\begin{enumerate}
    \item \textbf{Deployment}: The \textit{Owner} initiates the process by deploying three key contracts: \texttt{PricingLogic}, \texttt{ModelPointsLogic}, and \texttt{InsuranceLogic}. The initial parameters are set as follows:
    \[
    \eta = 0.1, \quad \alpha_{\text{SCR}} = 0.995, \quad \text{and} \quad \alpha_{\text{MCR}} = 0.85.
    \]

    \item \textbf{Initial Funding}: A \textit{Surplus Provider} kicks things off by sending 0.1 ETH to the smart contract, providing the initial capital needed to underwrite policies.

    \item \textbf{Policy Underwriting}: \textit{Policyholders} begin underwriting various insurance policies to protect against precipitation exceeding 5 units in Marseille or Strasbourg on specific days of the year. Here are the details of the policies they secure:
    \begin{itemize}
        \item Policy \#1: Covers Marseille on day 60, with a liability of $0.01$ ETH.
        \item Policy \#2: Covers Strasbourg on day 60, with a liability of $0.01$ ETH.
        \item Policy \#3: Covers Marseille on day 45, with a higher liability of $0.02$ ETH.
        \item Policy \#4: Covers Strasbourg on day 80, with a liability of $0.005$ ETH.
        \item Policy \#5: Covers Strasbourg on day 102, with a liability of $0.015$ ETH.
        \item Policy \#6: Covers Strasbourg on day 206, with a liability of $0.01$ ETH.
        \item Policy \#7: Similar to Policy \#6, covering Strasbourg on day 206, with a liability of $0.01$ ETH.
        \item Policy \#8: Covers Marseille on day 300, with a liability of $0.015$ ETH.
        \item Policy \#9: Covers Marseille on day 282, with a liability of $0.005$ ETH.
        \item Policy \#10: Covers Marseille on day 180, with a liability of $0.02$ ETH.
    \end{itemize}

    \item \textbf{Settling one Contract and withdrawing of the investor}: The \textit{Owner} starts settling the contracts based on the observed precipitation:
    \begin{itemize}
        \item Contract \#3 is settled with precipitation in Marseille on day $45$ not exceeding the threshold.
        \item A \textit{Surplus Provider} decides to burn $0.035$ tokens, adjusting their investment.
    \end{itemize}

    \item \textbf{Further Contract Settlements}: The \textit{Owner} continues settling more contracts:
    \begin{itemize}
        \item Contract \#1 is settled with precipitation in Marseille on day $60$ exceeding the threshold.
        \item Contract \#2 is settled with precipitation in Strasbourg on day $60$ exceeding the threshold.
        \item Contract \#4 is settled with precipitation in Strasbourg on day $80$ exceeding the threshold.
        \item Contract \#5 is settled with precipitation in Strasbourg on day $102$ not exceeding the threshold.
    \end{itemize}
\end{enumerate}

The final event triggers the smart contract to file for bankruptcy, as the surplus \(X\) falls below the Minimum Capital Requirement (MCR) threshold. Consequently, the premiums for policies \(\#6\), \(\#7\), \(\#8\), \(\#9\), and \(\#10\) are refunded to the policyholders. The remaining surplus is then returned to the surplus provider.

We do not suggest that the participants in this scenario acted strategically—in fact, quite the opposite. By withdrawing funds early, the surplus provider jeopardized the solvency of the smart contract. The scenario where multiple insurance contracts result in compensation is both highly unlucky and statistically improbable. Additionally, the reduction in the number of model points below the minimum required for the Cornish-Fisher approximation led to a drastic increase in the MCR, which ultimately exceeded the surplus. These choices were intentional, designed to illustrate the consequences of the smart contract going bankrupt and undergoing a reset.

Deploying the contract results in a ``contract creation'' transaction on the blockchain. Each subsequent function call also generates a transaction, which is recorded on the blockchain. These transactions can be retrieved and organized into a data frame, the structure of which is detailed in \cref{tab:transactions_dataframe_dictionnary}.

\begin{table}[h!]
\footnotesize
\centering
\begin{tabular}{llll}
\toprule
\textbf{Variable Name} & \textbf{Type} & \textbf{Example Value} & \textbf{Description} \\
blockNumber & object & 8518222 & The height of the block in the blockchain \\
timeStamp & object & 1749558720 & The date and time when the transaction was created or mined \\
hash & object & 0xe4ca$\ldots$ & The unique hash value identifying the transaction \\
nonce & object & 101 & A number used to ensure that each transaction can only be processed once \\
blockHash & object & 0x0248$\ldots$ & The unique hash value identifying the block that contains the transaction \\
transactionIndex & object & 31 & The position of the transaction within the block \\
from & object & 0xe8e7$\ldots$ & The address of the sender who initiated the transaction \\
to & object & 0xbe03$\ldots$ & The address of the recipient of the transaction \\
value & object & 0 & The amount of ETH sent with the transaction \\
gas & object & 3958691 & The maximum amount of gas allocated for the transaction \\
gasPrice & object & 1501121993 & The price of each unit of gas, specified in ETH \\
isError & object & 0 & Indicates whether an error occurred during the transaction \\
txreceipt\_status & object & 1 & Indicates the status of the transaction receipt \\
input & object & 0x6080$\ldots$ & The data field containing information necessary for the transaction \\
contractAddress & object & 0xbe03$\ldots$ & The address of the contract created as a result of the transaction \\
cumulativeGasUsed & object & 6358722 & The total amount of gas used in the block up to and including this transaction \\
gasUsed & object & 3926362 & The amount of gas actually used by this specific transaction \\
confirmations & object & 652 & The number of blocks mined after the block containing this transaction \\
methodId & object & 0x6080$\ldots$ & The identifier for the method being called in a smart contract \\
functionName & object & & The name of the function called in a smart contract \\
\bottomrule
\end{tabular}
\caption{Description of the transactions data frame associated with the contract.}
\label{tab:transactions_dataframe_dictionnary}
\end{table}

\begin{figure}[!ht]
\centering
  \includegraphics[width=0.6\linewidth]{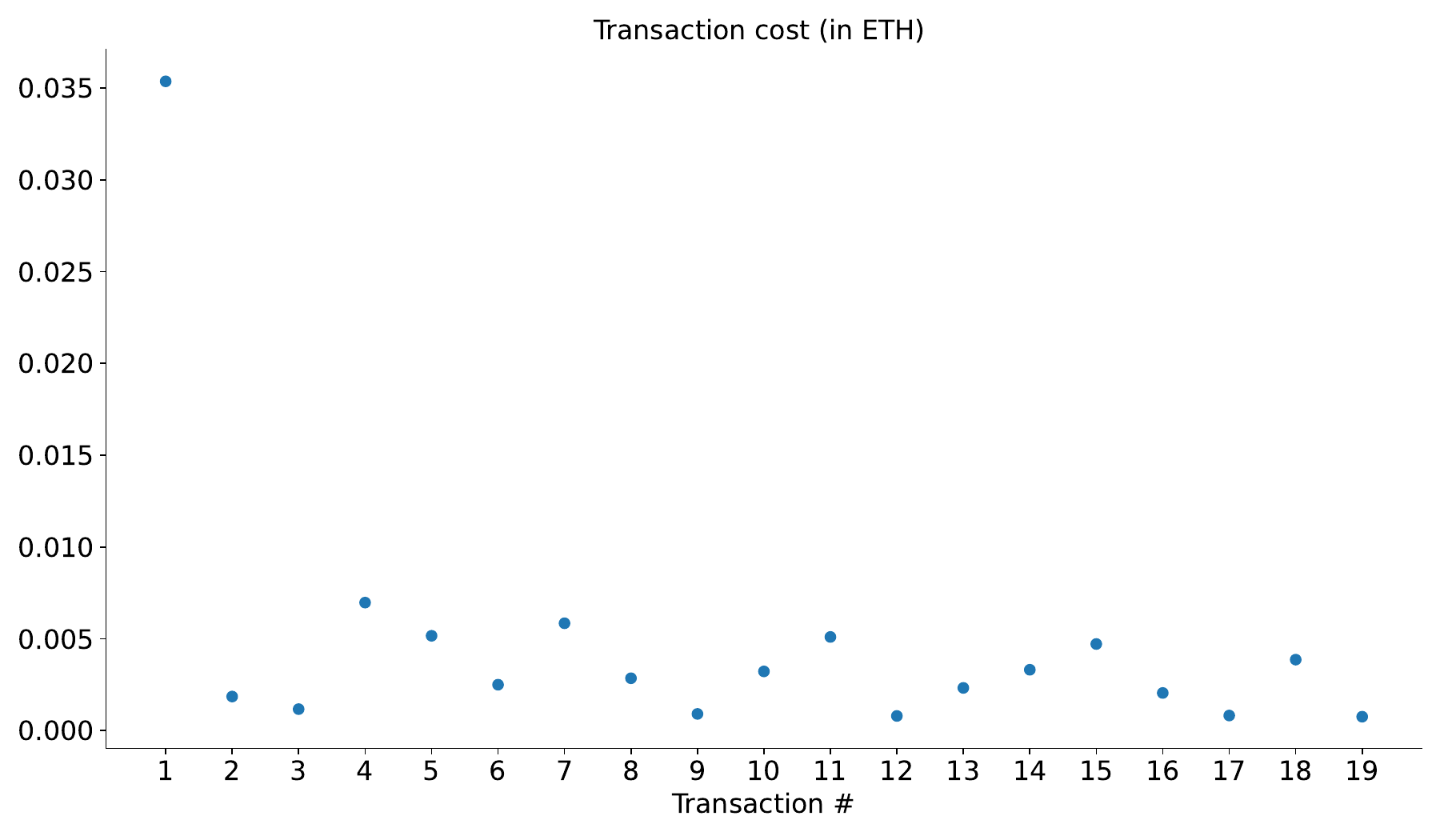}
  \caption{Transaction costs in ETH.}
  \label{fig:transaction_cost}
\end{figure}

As mentioned earlier, each transaction incurs a cost to the sender, calculated as:  
\[
\texttt{gasPrice} \times \texttt{gasUsed}.
\]
The value of a unit of gas (\texttt{gasPrice}) fluctuates over time, as it depends on the level of activity on the Ethereum network. High network congestion leads to higher gas prices. Another factor influencing transaction costs is the amount of data being recorded, which is reflected in the \texttt{input} field of the transaction details. The "contract creation" transaction is by far the most expensive, with a cost of approximately \(\text{ETH}0.035\), due to the significant amount of data required to deploy the smart contract. In contrast, subsequent transactions typically cost between \(\text{ETH}0.002\) and \(\text{ETH}0.008\), as illustrated in \cref{fig:transaction_cost}.

In addition to the list of all transactions, we can retrieve the log of events emitted by the smart contract. A common practice is to associate a specific event with each function. Function calls impact the state of the system, and events provide a mechanism to monitor state transitions effectively. However, care must be taken when defining events, as events that contain a large amount of information will require more data storage and, consequently, increase transaction costs. The definition of the events are provided in \cref{tab:events}.

\begin{table}[!ht]
\footnotesize
\centering
    \centering
    
    \begin{tabular}{|l|l|rrrp{3cm}|}
        \toprule
        \textbf{Event Name} & \textbf{Short Name} & \textbf{Logged Information} & \textbf{Type} & \textbf{Example Value} & \textbf{Description} \\ 
        \midrule
        \multirow{4}{*}{ParametersUpdated} 
        & \multirow{4}{*}{Update}
        & \texttt{newEta} & int & $1000$ & premium loading \\
        &&\texttt{newQAlphaSCR} & int & $25758$ & Quantile of the normal distribution\\
        &&\texttt{newQAlphaMCR} & int & $10364$ & -\\
        \midrule
        \multirow{3}{*}{Fund} 
        & \multirow{3}{*}{Fund} & from & hex &  $0x123\ldots$& address of the sender\\
        && x&int& 10000 & Amount of ETH \\
        &&y &int& 10000& Amount of token\\
        \midrule
        \multirow{3}{*}{Burn} 
        & \multirow{3}{*}{Burn} & from & hex &  $0x123\ldots$& address of the sender\\
        && x&int& 10000 & Amount of ETH \\
        &&y &int& 10000& Amount of token\\
        \midrule
        \multirow{5}{*}{InsuranceUnderwritten} & \multirow{5}{*}{Underwrite}& contractId&int&1&Index of the policy\\
        &&customer$\_$address&hex&$0x456\ldots$&Ethereum address\\
        && T&int&80& Day of the event with year 2025\\
        && station&string&STRASBOURG-ENTZHEIM& location of the event\\
        && l&int&1000000&Compensation if the event occurs\\
        && cp&int&1000000&Commercial premium\\
        && status&int&0&Status of the insurance policy \\
        && SCR&int&100000&Updated Solvency Capital Requirement \\
        && MCR&int&100000&Updated Minimum Capital Requirement \\
        \midrule
        \multirow{4}{*}{ClaimSettled}&\multirow{4}{*}{Settle}&contractId&int&1&Index of the policy\\
        &&customer$\_$address&hex&$0x456\ldots$&Ethereum address\\
        &&payoutTransferred&boolean& \texttt{True}& Indicates if a compensation has been paid\\
        && SCR&int&100000&Updated Solvency Capital Requirement \\
        && MCR&int&100000&Updated Minimum Capital Requirement \\
        \bottomrule
    \end{tabular}
    \caption{Description of Events Generated by the Smart Contract}
    \label{tab:events}
\end{table}

\begin{figure}[!ht]
\centering
  \includegraphics[width=0.6\linewidth]{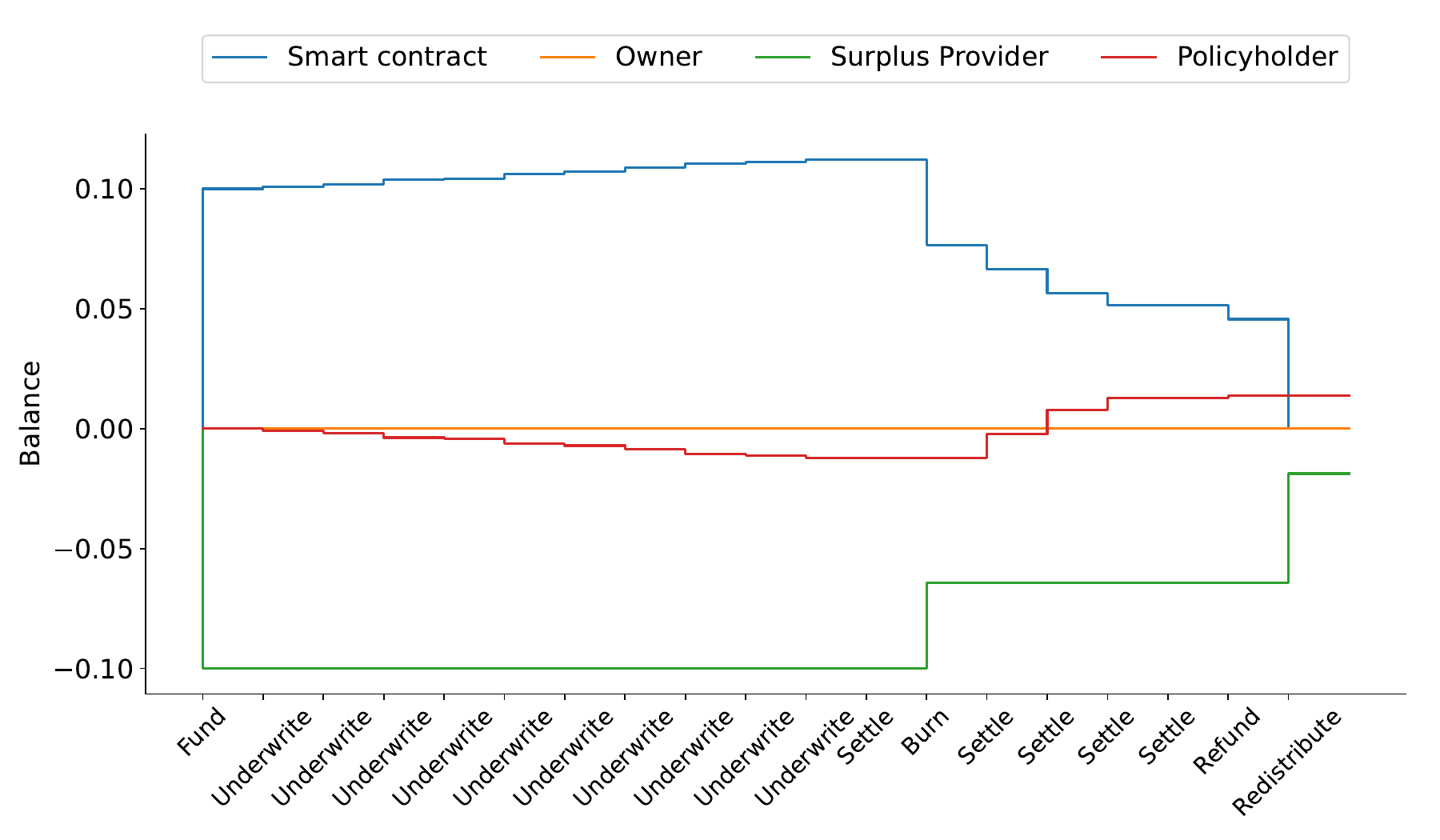}
  \caption{Balance in ETH of the Ethereum address of the \textit{Owner}, the \textit{Surplus Provider}, the \textit{policyholder}, and the \textit{Smart Contract}.}
  \label{fig:balance}
\end{figure}

The events used in this study are defined in the Solidity code, which is included in the online supplementary materials accompanying this article. Using Python, we retrieve the event logs to track participant balances over time. \Cref{fig:balance} illustrates the balances, in ETH and before transaction fees, of the Ethereum addresses associated with the \textit{Owner}, the \textit{Surplus Provider}, the \textit{Policyholder}, and the \textit{Smart Contract} after each function call.

\begin{figure}[!ht]
\centering
  \includegraphics[width=0.6\linewidth]{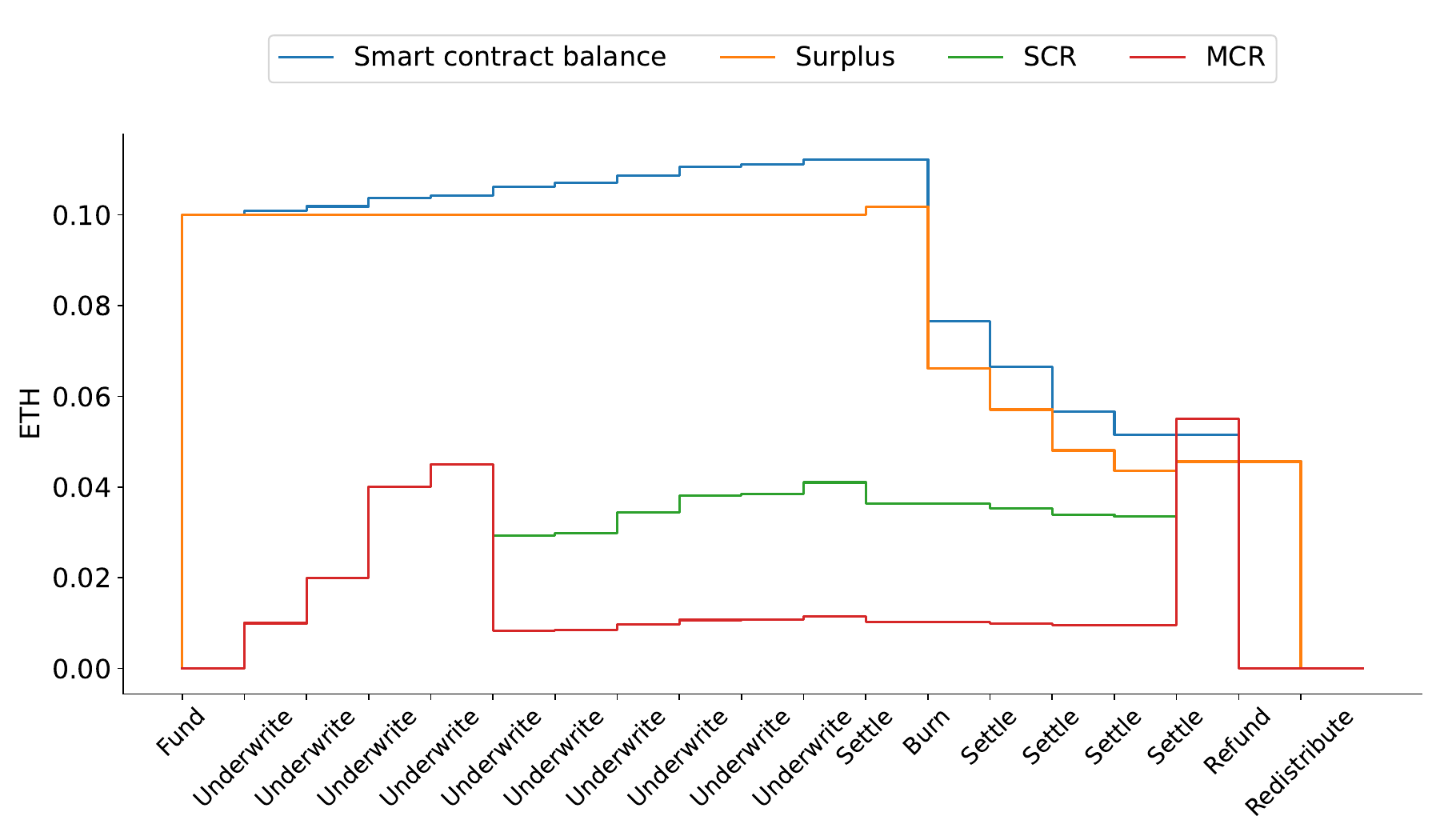}
  \caption{Trajectories of $X$, $B$, $\text{SCR}$ and $\text{MCR}$.}
  \label{fig:surplus}
\end{figure}

We observe that the scenario was particularly beneficial to the \textit{Policyholder}, who received three compensation payments and a refund for five policies. The event logs also enable us to visualize the trajectories of the various processes defined in \cref{sec:parametric_insurance} and \cref{sec:smart_conract_for_parametric_insurance}. Specifically, \cref{fig:surplus} displays the trajectories of \(X\) (surplus), \(B\) (balance), \(\text{SCR}\) (Solvency Capital Requirement), and \(\text{MCR}\) (Minimum Capital Requirement) as they evolved throughout the scenario.

Note the difference between the surplus and the balance of the smart contract: the balance of the smart contract increases at each "Underwrite" event, while the surplus remains constant. Until contract $\#5$ is underwritten, the SCR and MCR are exactly equal to the sum of all potential compensations to be paid. Once contract $\#5$ is underwritten, the number of model points reaches $5$, and the solvency capitals are calculated via the Cornish-Fisher approximation, causing a drop in the value of the solvency capitals. After the settlement of contract $\#3$, the exchange rate of protocol tokens against ETH appreciates, which is why the surplus provider decides to cash out part of her investment. The compensations paid after the settlements of contracts $\#1$, $\#2$, and $\#4$, combined with the fact that the settlement of contract $\#5$ reduces the number of model points below $5$, cause the smart contract to reset as the surplus becomes smaller than the MCR. The status of all active policies changes to "cancelled." Premiums are refunded to the policyholders, and investors receive any remaining balance from the smart contract.

\begin{remark}\label{rem:drop_solvency_capital_explanation}
The drop in value of both the MCR and the SCR after including contract $\#$5 might seem surprising at first glance. Under Solvency II regulations, there is an absolute floor on the MCR value, set at least to {\euro}2.7 million euros for non-life insurance companies. We have proposed an alternative method for defining this floor, based on the sum of exposures when the number of independent risks is too small (fewer than 5 in \cref{fig:surplus} for illustration purposes, but typically fewer than hundreds or thousands in real-life scenarios). In practice, a decrease in solvency capital can occur during the transition from the standard formula to internal models, which somewhat corresponds to adopting normal approximations in our example.
\end{remark}
The Solidity code for the smart contract and the Python scripts for analyzing blockchain data are freely available in the \href{https://github.com/LaGauffre/smart_parametric_insurance}{smart$\_$parametric$\_$insurance} GitHub repository.

\section{Limitations and perspectives}\label{sec:limitations}
This article proposes a framework for a smart contract suited to parametric insurance and demonstrates the feasibility of its implementation on a blockchain. Additionally, it makes this open-source prototype available to other researchers, facilitating further exploration of its limitations and potential enhancements. In this section, we outline the main limitations of this initial approach and suggest possible solutions.




The underlying risks are considered stationary. However, in practice, many risks suitable for smart contracts with parametric triggers are influenced by regime changes, which can lead to event clustering or increased claim frequency. To mitigate adverse selection, a critical first step in risk management is introducing a waiting period before the guarantee becomes active. This measure helps prevent policyholders from exploiting arbitrage opportunities when they are aware of worsening risks. For instance, in the case of flight disruptions, events like strikes, conflicts, or volcanic eruptions can make delays almost certain in the short term. Similarly, for weather-related risks, extreme event warnings may be issued days in advance. It is essential to ensure that policyholders cannot exploit the smart contract under such circumstances while paying the same premium as usual.

In some cases, a waiting period may not be sufficient to mitigate risk, as underlying risks can be anticipated several months in advance. For example, the El Ni\~{n}o/La Ni\~{n}a cycle, which is partially predictable months ahead, increases the likelihood of certain extreme weather events while reducing others. Moreover, long-term factors such as climate change and sectoral inflation can exacerbate risks over time. If competitors actively adjust their premiums to reflect these changes while the smart contract does not, adverse selection becomes a likely outcome. Finding a way to update premiums dynamically is essential. This could be achieved either automatically—by scraping new data from competitors, insurance federations, or other relevant sources—or through decisions made by a "board" of customers and investors, who could leverage voting rights granted via tokens. A governance system where token-holding investors form a board with voting rights aligns with the concept of decentralized organizations, which are extensively studied in management science and have naturally emerged within blockchain applications. However, classical issues such as over- or under-reactions stemming from individual risk perceptions and cognitive biases could be amplified, especially if voter participation is low or voting power is concentrated due to coalitions. These challenges are not uncommon, as low or uneven participation rates have frequently been observed in various decentralized networks, we refer the reader to the work of \citet{messias2023understanding} for a comprehensive discussion.





Our smart contract is exposed to the risk of a "mass investor lapse," where investors may collectively withdraw their tokens. Stronger incentive mechanisms could be implemented to encourage investors to retain their tokens, particularly when the contract's balance approaches the risk capital threshold. One potential approach is to introduce a coin age principle, where tokens held for longer periods gain additional value. This concept has already been explored in the implementation of the cryptocurrency PeerCoin\footnote{See \url{https://www.peercoin.net/}}. A traditional approach to managing the surplus of an insurance company is to implement a dividend policy that rewards investors. This can be achieved using classical optimal dividend strategies. Barrier or band strategies, for instance, have been shown to be optimal across various modeling frameworks (see the survey by \citet{avanzi2009strategies}). In our framework, we operate within a discrete-time risk model with capital injections and withdrawals. This approach is consistent with the structure of our system, where wealth is only inspected at discrete points—namely, during settlements or when an investor funds or withdraws from the contract.

In this paper, we have not addressed regulatory risks. In many countries, a smart contract like the one proposed would need to comply with insurance regulations. For instance, in the European Union, risk governance would need to be developed to align with the requirements of Pillar 2 of Solvency II, which presents significant challenges. Additionally, adhering to anti-money laundering (AML) regulations may necessitate incorporating a KYC (know your customer) module. While implementing such a module can be particularly challenging in a blockchain-based environment, some cryptocurrency wallet startups have successfully tackled this issue, providing potential avenues for further development.

In our approach, surplus management is handled through proxies referred to as the Minimum Capital Requirement (MCR) and Solvency Capital Requirement (SCR). If such a smart contract were to be classified as an insurance product, it would need to comply with the absolute minimum MCR threshold prescribed by Solvency II in Europe. 

Exchange rate risk is a significant concern for potential customers, particularly as compensations are paid in ETH, a cryptocurrency subject to high volatility against fiat currencies. To mitigate this issue, one potential solution is to link our contract with another smart contract designed to exchange ETH for a stablecoin. This approach leverages the interoperability of blockchain applications, a key advantage of decentralized finance over traditional financial systems. 

If the contract is not classified as an insurance product, it could lead to tax and regulatory implications. To ensure recognition as insurance, it is crucial to keep basis risk sufficiently low. Effective basis risk management not only protects policyholders but also mitigates the risk of the contract being reclassified as gambling. In some countries, policyholders may be required to provide a brief proof of loss to validate claims. While we assume that the occurrence of a compensation-triggering event inherently results in a loss for the policyholder, real-world scenarios might necessitate additional efforts from either the policyholder or the smart contract’s governance system to address such requirements. Basis risk management is a central problem when dealing with parametric insurance solution and as such under investigation in many recent research work like that of \citet{Niakh2025}.

Sustainability considerations also warrant further investigation. While blockchain technology consumes significant amounts of electricity and water, its environmental impact could be partially offset by reducing the carbon emissions associated with traditional insurance operations, such as staffing, office infrastructure, and claim management processes. 

In conclusion, we have developed a modest beta version of a smart contract for parametric insurance on the Ethereum blockchain. The Solidity code for the smart contract and the Python scripts for analyzing blockchain data are freely available in the \href{https://github.com/LaGauffre/smart_parametric_insurance}{smart$\_$parametric$\_$insurance} GitHub repository. Beyond serving as a proof of concept, this work identifies several research questions that need to be addressed to make the product fully operational. We hope that our open-source beta version will provide a foundation for other researchers to build upon, facilitating the development of enhanced features and overcoming the limitations we have outlined.

\section*{Acknowledgements}
Pierre-Olivier Goffard's work is conducted as part of the Research Chair DIALOG\footnote{See \url{https://chaire-dialog.fr/en/welcome/}}, under the auspices of the Risk Foundation, an initiative by CNP Assurances. His research is also supported by the ANR project BLOCKFI\footnote{See \url{https://anr.fr/Project-ANR-24-CE38-7885}}. Stéphane Loisel acknowledges support from the Research Chair ACTIONS\footnote{See \url{https://chaireactions.fr/}} (funded by BNP Paribas Cardif), the research initiative Sustainable Actuarial Science and Climatic Risks\footnote{See \url{https://sites.google.com/view/stephaneloisel/recherche/projets/actuariat-durable-et-risques-climatiques}} (funded by Milliman Paris), and the ANR project DREAMES\footnote{See \url{https://anr.fr/Projet-ANR-21-CE46-0002}}.
\appendix

\end{document}